# Social Media, Money, and Politics:
# Campaign Finance in the 2016 US Congressional Cycle

Lily McElwee[1] and Taha Yasseri[1,2,*]

[1]Oxford Internet Institute, University of Oxford, Oxford, UK
[2]Alan Turing Institute, London, UK

*Correspondence:
Taha Yasseri
taha.yasseri@oii.ox.ac.uk

**Abstract**
With social media penetration deepening among both citizens and political figures, there is a pressing need to understand whether and how political use of major platforms is electorally influential. Particularly, the literature focused on campaign usage is thin and often describe the engagement strategies of politicians or attempt to quantify the impact of social media engagement on political learning, participation, or voting. Few have considered implications for campaign fundraising despite its recognized importance in American politics. This paper is the first to quantify a financial payoff for social media campaigning. Drawing on candidate-level data from Facebook and Twitter, Google Trends, Wikipedia page views, and Federal Election Commission (FEC) donation records, we analyze the relationship between the topic and volume of social media content and campaign funds received by all 108 candidates in the 2016 US Senate general elections. By applying an unsupervised learning approach to identify themes in candidate content across the platforms, we find that more frequent posting overall and of issue-related content are associated with higher donation income when controlling for incumbency, state population, and information-seeking about a candidate, though campaigning-related content has a stronger effect than the latter when the number rather than value of donations is considered.

# Introduction

Scholars have sought to understand the relationship between technology and democracy since the 1990s (Barber, 1998). With rapidly rising adoption of social media by citizens, US politicians are increasingly aware of the power of major platforms to communicate and organize for political purposes (Gainous and Wagner, 2014; Margetts, John, Hale, & Yasseri, 2015). The growth in social media campaigning specifically has been mirrored by growth in literature analyzing usage itself and implications for a range of electorally-related outcomes (Boulianne, 2015). As social media penetration continues to deepen among the American electorate, there is pressing need to determine whether and how political candidates' use of these platforms has electoral significance.

Social media adoption is widespread for any protocol, with 70% of US adults on Facebook and 20% on Twitter as of 2016. Empirical work has demonstrated a rise in citizen and candidate interaction on the main platforms. Roughly 40% of Americans had posted and 80% had seen political content on social networking sites (SNSs) as of April 2014. Most importantly, followership of political figures on the main SNSs is on the rise; while 14% and 24% of 18-29 year olds and 6% and 21% of 30-49 year olds followed elected officials, political parties, or candidates for office in 2010 and 2014 respectively (Anderson, 2015), 35% of the American online population now does so (Kalogeropoulos, 2017). The practice of followership is bipartisan, with supporters of both parties equally likely to follow political figures on social media (Anderson, 2015).

Studies seeking to explain increased campaign usage distill unique offerings of major social media platforms. Most, primarily focused on Twitter, argue such sites facilitate the "most inexpensive, unmediated, and closely focused forms of communication in campaign history," (Gainous and Wagner, 2014, 54); further, these platforms are ideally suited to the types of messaging in which office-seekers want to engage, as they enable candidates to create succinct themes and highlight victories rather than explain the minutiae of complex legislation. It has been suggested that although the major social media platforms were not originally created for political purposes, the fact that they are low cost, allow direct communication with the public, and provide access to a wide body represent advantages over traditional phone, mail, and website-based campaigning (Auter and Fine, 2017). Because social media has been shown to be fundamentally different from 'campaigning as usual' (Bode et al, 2016), the implications of rising use of social media in campaigning are worth investigation.

Cognizant that social media are increasingly commonplace, a plethora of studies have begun to describe usage and test impact across a range of electoral areas, such as vote gains (Yasseri & Bright, 2016; Bright et al., 2017), political participation (Bode, 2012), and political learning (Dimitrova et al., 2014). These studies offer a wealth of information regarding the ways in which online campaigning is playing a role in electoral processes, but leave addressable gaps. Those examining implications often focus on volume rather than content of social media activity as the explanatory variable. Those classifying political social media content categorically fail to examine broader electoral implications of technologies.

The past literature on social media electoral campaigning examines determinants of use (Jackson and Lilleker, 2011), genres of content (Bode, 2016; Gainous and Wagner, 2014), and implications for electorally-related outcomes (Bright et al., 2017). Quantifying the effect of social media on electoral outcomes is a work in progress, but initial analyses suggest significance, contradicting skepticism from media and academic accounts (Bright et al, 2017). The field will likely continue to grow in sync with political enthusiasm for social media platforms.

On the other hand, the wide body of studies classifying social media posts by type or topic or assessing determinants behind specific styles fail to examine the influence of such classifications on various elements of electoral success. Bode et al., (2016) analyze all 10,303 tweets by campaigns in the 2010 senatorial elections, classifying each according to seven topics (economic, social issues, foreign policy, social welfare, law and order, environment/energy, other); while they assess the relationship between social media activity and campaign resources and competitiveness, they focus on tweeting volume rather than topics and do so in a bid to predict the former rather than the latter. Evans, Cordova, and Sipole (2014) examine how candidates for the House of Representatives used Twitter during the 2012 cycle, classifying tweets according to a six-part scheme. While they find a clear distinction between the tweeting styles of incumbents, Democrats, women, and those in competitive races versus challengers, Republicans, minor party candidates, men, and those in safe districts, they do not examine electoral implications of these stylistic differences. Jones, Noorbalachi et al., (2016) similarly consider the demographic breakdown of specific types of content, finding that Republican and conservative legislators stress values of tradition, conformity, and national security, whereas Democratic and liberal legislators stress values of benevolence, hedonism, universalism, and social/economic security. However, they leave investigation at the content level, rather than evaluate relevance to future electoral chances.

Gainous and Wagner (2014) offer perhaps the most comprehensive and theoretically-grounded analysis of candidate activity on Twitter, creating a four-part typology of political Twitter content based on research into the patterns and activities of modern campaigning. They research determinants of social media adoption and use, analyzing both total tweet volume and types of Twitter use for each of the four types of online campaigning in light of a variety of political and demographic factors. Despite identifying differences across partisan, incumbency status, congressional office (House versus Senate) and gender metrics through bivariate, multivariate, and qualitative examination, they do not test the association between these typological breakdowns and electoral outcomes of any sort.

In one of the first large-scale empirical studies linking vote outcomes to Twitter use, Bright et al (2017) comprehensively account for various elements of candidate social media activity in the 2015 UK general election, such as volume of posting, followership, and dialogue with followers, but do not incorporate specific topics of content in assessing effects. Others lump candidate posts with candidate-related posts (Murphy, 2015) or rely on mentions of political candidates and political parties (Tumasjan et al., 2010).

Money is an important ingredient in US congressional elections. General election candidates in the 2016 Senate race alone raised $667,697,881, with Democrats

outstripping Republicans at $363,396,637 to $302,100,403.[1] In contrast to many other liberal democracies, private funding in the United States is often a primary campaign resource. Since 1971, such funding has been governed by the Federal Election Campaign Act, amendments to which cap individual donations to $2,700 per election and mandate disclosure for all contributions received above $200. Registered political committees, such as candidate campaigns, file reports with the Federal Election Commission (FEC), which are made publicly-available within 48 hours of submission but updated continuously afterwards as the numbers become more accurate. These requirements make the role of money in US elections both influential and researchable.

There has been surprisingly little attention to the impact of either volume or type of usage on campaign fundraising success. To the best of our knowledge, there are only few studies analyzing social media effects on political donations (Hong, 2013; Petrova, Sen, & Yildirum, 2016), but both of these works use politicians rather than candidates as the unit of analysis and focus solely on adoption, rather than content type, as the explanatory variable.

Campaign contributions rely on decisions about whether and how much to give. Studies in political behavior suggest patterns in such decisions, finding that 1) individuals use social information, or information about what others are doing or have done, to select whether and how much to contribute (Margetts et al., 2011; Croson & Shang, 2008; Traag, 2016); 2) individuals require information about which candidates represent their political beliefs in order to make contributions (Grant and Rudolph, 2002); and 3) selective targeting has the potential to temper the impact of income on contribution decisions, such that campaigns can maximize donations by constructing a messaging and solicitation strategy that aligns with the background information of supporters and their associates.

In this work, we analyze this relationship between online campaigns and the donation received by drawing on publicly-available data from Facebook and Twitter timelines, Google Trends, Wikipedia page views, and FEC donation records for all 108 candidates in the 34th general elections for US Senate in 2016.

## Data and Methods

We collect data on the number and sum of donations received and the social media activity conducted by general election senatorial candidates during the six-week period prior to Election Day in the 2016 cycle. For details see Supplementary Information.

***Donation records***. The FEC provides searchable donation records based on reports from registered campaign committees. We use MapLight to retrieve the total sum and count of donations received by the candidates with campaign accounts.[2] The customized dataset includes 83 of 108 candidates. Twenty-five candidates are missing donation information altogether across the five periods (see Supplementary Information for a full list). Candidates are not required to file with the FEC if they receive or spend under $5000, so the limited totals of these candidates is the most likely explanation for absent records, especially since data collection falls sufficiently after reporting deadlines.

---
[1] Statistics retrieved from the Center of Responsive Politics.
https://www.opensecrets.org/overview/index.php
[2] https://maplight.org/

*Social media*. We use raw Facebook posts and tweets as a basis for the social media variables. We manually search within each social media platform for candidate handles and check for additional accounts via Google. Many candidates have more than one Twitter or Facebook account. For each candidate, we include all accounts found on a given platform to capture the entirety of his or her Twitter or Facebook presence.

*Google Trends*. We collect daily data on Google search trends over the past year for each candidate's full name via the gtrendsR package in R.[3]

*Wikipedia*. We manually search for candidates' Wikipedia entries; of 108 candidates, 71 have dedicated Wikipedia articles. The list provides a basis for collection of daily data on Wikipedia page views for each available article, which we do through the wikipediaR package (Bar-Hen et al., 2016).[4]

*Incumbency*. We collect information on whether a candidate is an incumbent from Ballotpedia.[5]

*Population*. For each candidate, we collect data on the population of their state from 2016 census figures. This control variable has been reported to play a role in campaign finance (Lin, Kennedy & Lazer, 2017).

*Type of post*. To classify posts by topic, we use latent Dirichlet allocation (LDA) topic modeling, which postulates a latent structure in a corpus and represents each document as a distribution over topics (Steyvers and Griffiths, 2007). Adopting topic modeling identifies topics of content, rather than categories of content, thereby distinguishing any output from existing categorizations of political content online discussed above (e.g., Gainous and Wagner, 2014; Evans et al, 2014).

For the details on corpus cleaning and optimization of the topic modelling algorithm, see Supplementary Information.

*Classifying Documents*. The topic modelling process results in ten collections of words (Figure 1). Consistent with precedent (Messing et al., 2014), topic labels are selected according to the words appearing most frequently in the posts classified by a specific topic (see Supplementary Information).

---

[3] https://cran.r-project.org/web/packages/gtrendsR/gtrendsR.pdf.
[4] https://cran.r-project.org/web/packages/WikipediaR/WikipediaR.pdf.
[5] https://ballotpedia.org/United_States_Senate_elections,_2016.

| Topic | Name & Description |
|---|---|
| 1 | Campaigning metatalk, related to meetings, calls, and stops. |
| 2 | Policy issues, mostly related to education ('school', 'student', 'educ', 'learn') and security ('secur'). |
| 3 | Campaign topics, related to voting/election. |
| 4 | Campaign topics, related to partisan talk and discussion of federal topics, e.g. Democrat/Republican ('donald' and 'clinton'). |
| 5 | Policy issues, most related to veterans ('veteran', 'famili') and taxation ('tax') |
| 6 | Policy issues, mostly related to healthcare ('care', 'health', 'afford') but also others. |
| 7 | Policy issues, related to community ('leader', 'discuss', 'together', 'issu') and education ('colleg') |
| 8 | Campaign topics, such as endorsements and media appearances |
| 9 | Campaign topics, related to events ('time', 'event', 'tomorrow') and solicitation of volunteers. |
| 10 | Policy issues, mostly related to jobs ('job', 'worker', 'protect') |

**Figure 1. Topic Modeling Output: Qualitative Description**. *Brief description of the ten topics discovered through modeling, see Supplementary Information for more details.*

*Grouping Topics.* Analysing similarity between topics provides a basis for grouping individual topics and deriving a meaningful hypothesis about the relationship between fundraising and specific types of content. We use cosine similarity to identify the similarity of the automatically-generated topics in the LDA model, an approach with precedent in thematic topic aggregation (Messing et al, 2014).

Setting the weights for all pairs with similarity smaller than 0.23 to zero, we visualize the topic similarity for those pairs above the cutoff in graphing platform Gephi.[6] Analyzing modularity via the Louvain community detection algorithm (Blondel, Guillaume, Lambiotte, and Lefebvre, 2008), we detect communities of topics that are more similar to each other than to the others (Figure 2).

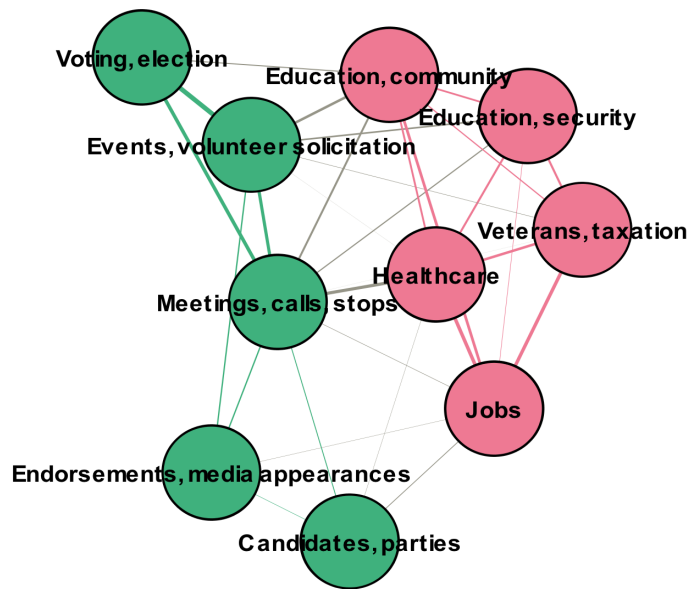

**Figure 2. Topic Network Graph**. *Based on cosine similarity weightings, the Louvain community detection method discovers two communities of topics, which qualitative examination shows to group thematically into campaigning-related (green) and issue-related (pink) posts.*

---

[6] https://gephi.org/.

It becomes evident that there are two main clusters of topics. Topics 1, 3, 4, 8, 9 form a community while topics 2, 5, 6, 7, 10 form a separate community. The former group all have campaigning-related elements while the other topics relate to policy issues. In the following analysis, we consider the social media posts in each group separately.

## Results

*Overall Description*

The total number of donations per candidate, per week ranges from 1 to 8131 (mean 698 and median 108). The total sum received per candidate, per week ranges from $19 to $1,600,000 (mean $178,200 and median $62,580). The distributions of both specifications are fat-tailed. The distributions are shown in Supplementary Information.

Figure 3 shows the relation between weekly donation counts and sum received. While the relationship does not appear linear across the count values, mean donation size falls with higher number of donations. This shows that some senatorial candidates receive a high number of smaller donations in certain periods, similar to Obama in the 2008 and 2012 campaigns (Bimber, 2014). The maximum of sum/count appears at roughly nine donations per week.

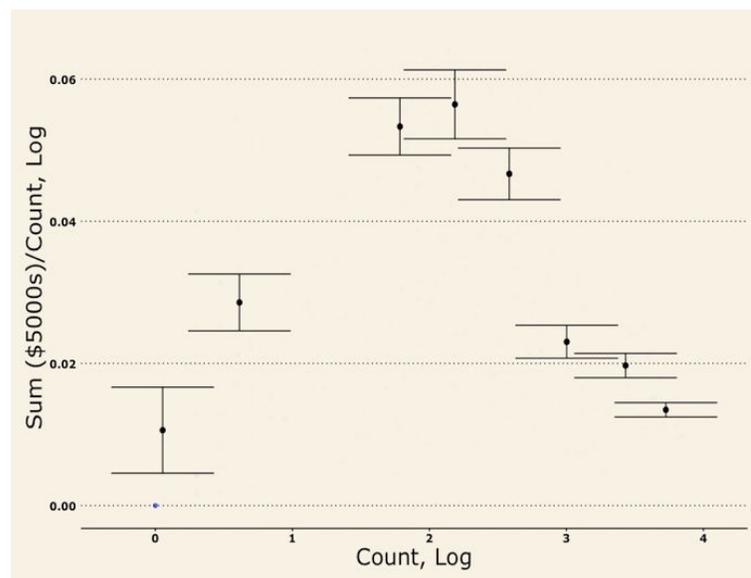

**Figures 3. Donation Count versus Size: Log/Log.** *Log transformation reveals a nonlinear non-monotonic relationship between the number and size of donations.*

Senatorial candidates tweeted on average much more frequently than they posted on Facebook. The weekly posts per candidate on Twitter range from 1 to 264 (mean 37 and median 24). The weekly posts per candidate on Facebook range from 1 to 37 (mean 8 and median of 5). The distributions are shown in Supplementary Information.

As Figure 4 shows, post volume for all topics detected by topic modelling rises as Election Day nears. In all periods, topic 1 (campaigning talk, including references to meetings, calls, and stops) is highly popular while topics 7 (issue-related, with reference to community) and 9 (campaign topics such as solicitation of volunteers) are relatively unpopular.

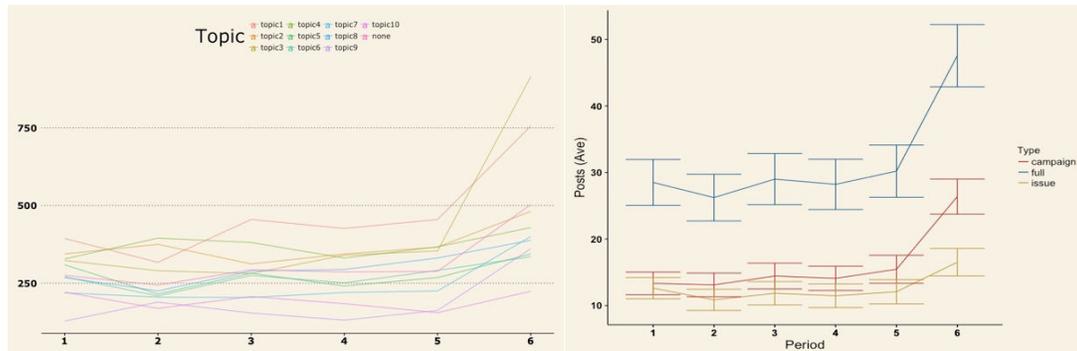

**Figure 4. Posting Volume Per Topic and Type, By Week**. *Posting across all topics rises as the Election nears.*

Weekly Google search volume for 2016 senatorial candidates runs from 0 to 1,360,791 (mean 25,870 and median 4,466). And Wikipedia weekly page views for those candidates with an article run from 0 to 1,360,791 (mean of 25,870 and median 4,466). The distributions are shown in Supplementary Information.

*Regression Analysis*

Simple scatterplots and residual plots reveal that the relationships between donations and both volume and type of post are not linear (see Supplementary Information). Considering this and the fact that the distributions of each variable are extremely fat-tailed, as demonstrated in the previous section, we transform all variables logarithmically. Also to be able to infer causality, we associate the values of donations (count and sum) in each week with the social media activity during the week before.

The linear regression model on the variables (Figure 5) show a positive and significant relationship between post volume and both donation totals and counts. This relationship is robust when controlling for state population and incumbency status as well as general information-seeking regarding the candidate, as proxied by Google Search Trends and Wikipedia page views. In the baseline model for donation sums, a 10% increase in post volume is associated with a 7% rise in donation income. When state population, incumbency, and general awareness are controlled for, the relationship weakens, with a 10% rise in post volume associated with a 1% boost in donation income. In terms of counts, for every 10% increase in post volume, the number of donations in the following week rises by 12%. When state population, incumbency, and general awareness are controlled for, this falls to a 3% boost.

Now we examine the effect of post type on donations, with the total posts as an additional control in order to account for the candidate's overall level of posting (Figure 6).

Similar to the post volume overall, campaigning-related posts have a positive and significant effect on both donation sums and counts with all controls in place. This positive relationship is significant when information-seeking is added as a control.

A 10% rise in campaign-related posts gives rise to a 4% rise in donation income in the multi-control model and a 10% rise in such posts produces an 8% boost in donation counts in the following period.

Now we turn to the issue-related posts (Figure 7). In a full-control model, a 10% rise in issue-related posts is associated with a 5% rise in both the donation income and counts.

It is notable that while the campaign-related posts are more effective in increasing the count of donations than the sum, the issue-related posts have similar effects on both the number of donations and their sum.

## Conclusion

Social media campaigning has normalized as a way for US political candidates of all backgrounds to promote their messages. 2017 marks near complete penetration of major social media platforms Twitter and Facebook by the 'political class'. Spurred by low barriers to entry, the march toward adoption by congressional candidates represents uniformity unique in the context of campaigning tools, simultaneously precluding any immediate conclusions about electoral implications and making these effects so pressing to study.

This study extends existing approaches by examining the effects of different content categories for an important component of elections. Methodologically, it builds on both these literatures by focusing on a novel output variable (donations), adopting a computational approach that avoids preconceived notions of post material, using candidates rather than politicians as a unit of analysis, and examining both Facebook and Twitter in assessing a candidate's social media presence.

Through these extensions, we offer both evidence of thematic patterns in social media electoral campaigning and a further step toward quantification of electoral payoffs through such campaigning. Topic modeling of general election senatorial candidates in the 2016 cycle reveals that posts discuss campaigning-related items or issue-related items, confirming that past classification methods based on more manual approaches indeed capture the spectrum of topics touched on by campaigning political figures online and hinting that this typology of candidate content is applicable up and down the US political ballot.

Regarding the effects of candidates' social media behavior, conclusions hold practical implications in the realm of modern campaigning. First, they point to a clear utility of investing resources in social media as a means to acquire additional campaign resources. Posting more frequently on social media appears to facilitate fundraising. Second, findings hint that some types of content may be associated with a higher return on posting than others. While both campaign- and issue-related posts are positively associated with donations, their effects are roughly equivalent on donation sums when incumbency, population, and information-seeking behavior are controlled for, and campaign-related posts have a stronger effect on the total number of donation receipts.

```
=================================================================================================
                                          Dependent variable:
                        -------------------------------------------------------------------------
                                          Donations (Log, 5000s)
                              Baseline          Partial Controls            Full
                                (1)                   (2)                    (3)
-------------------------------------------------------------------------------------------------
Post Volume                   0.701***              0.580***               0.144***
                              (0.036)               (0.041)                (0.040)

Incumbency                                          0.849***               -0.028
                                                    (0.138)                (0.117)

Population                                          0.153**                0.043
                                                    (0.062)                (0.050)

Google Search Trends                                                       0.057**
                                                                           (0.023)

Wikipedia Page Views                                                       0.288***
                                                                           (0.018)

Constant                      -0.153                -2.434***              -1.259*
                              (0.106)               (0.925)                (0.724)

-------------------------------------------------------------------------------------------------
Observations                   540                    540                    540
R2                             0.407                  0.448                  0.666
Adjusted R2                    0.406                  0.445                  0.663
Residual Std. Error      1.355 (df = 538)       1.310 (df = 536)       1.021 (df = 534)
F Statistic          369.842*** (df = 1; 538) 145.222*** (df = 3; 536) 213.231*** (df = 5; 534)
=================================================================================================
Note:                                                                *p<0.1; **p<0.05; ***p<0.01
=================================================================================================
                                          Dependent variable:
                        -------------------------------------------------------------------------
                                           Donations (Count)
                              Baseline          Partial Controls            Full
                                (1)                   (2)                    (3)
-------------------------------------------------------------------------------------------------
Post Volume                   1.214***              1.077***               0.315***
                              (0.057)               (0.065)                (0.058)

Incumbency                                          1.027***               -0.510***
                                                    (0.219)                (0.171)

Population                                          0.138                  -0.070
                                                    (0.098)                (0.073)

Google Search Trends                                                       0.126***
                                                                           (0.033)

Wikipedia Page Views                                                       0.497***
                                                                           (0.026)

Constant                      -0.113                -2.176                 -0.068
                              (0.165)               (1.463)                (1.055)

-------------------------------------------------------------------------------------------------
Observations                   540                    540                    540
R2                             0.459                  0.481                  0.734
Adjusted R2                    0.458                  0.478                  0.731
Residual Std. Error      2.112 (df = 538)       2.073 (df = 536)       1.487 (df = 534)
F Statistic          456.575*** (df = 1; 538) 165.459*** (df = 3; 536) 294.576*** (df = 5; 534)
=================================================================================================
Note:                                                                *p<0.1; **p<0.05; ***p<0.01
```

**Figure 5. Volume Model: Post Volume and Political Donations (Sum and Count)**

```
===============================================================================
                                   Dependent variable:
                  -------------------------------------------------------------
                              Donations (Log, 5000s)
                      Baseline       Partial Controls            Full
                        (1)                 (2)                   (3)
-------------------------------------------------------------------------------
Campaign-Related Volume  -0.186             0.256               0.383***
                        (0.174)            (0.185)              (0.144)

Post Volume            0.858***            0.357**              -0.195
                        (0.151)            (0.167)              (0.134)

Incumbency                                 0.934***              0.092
                                           (0.151)              (0.125)

Population                                 0.152**               0.043
                                           (0.062)              (0.050)

Google Search Trends                                            0.052**
                                                                (0.023)

Wikipedia Page Views                                            0.293***
                                                                (0.018)

Constant               -0.180*             -2.374**             -1.172
                        (0.109)            (0.925)              (0.721)

-------------------------------------------------------------------------------
Observations             540                 540                  540
R2                      0.409               0.450                0.671
Adjusted R2             0.406               0.446                0.667
Residual Std. Error  1.355 (df = 537)   1.309 (df = 535)    1.015 (df = 533)
F Statistic       185.540*** (df = 2; 537) 109.578*** (df = 4; 535) 180.863*** (df = 6; 533)
===============================================================================
Note:                                              *p<0.1; **p<0.05; ***p<0.01
```

```
===============================================================================
                                   Dependent variable:
                  -------------------------------------------------------------
                                   Donations (Count)
                      Baseline       Partial Controls            Full
                        (1)                 (2)                   (3)
-------------------------------------------------------------------------------
Campaign-Related Volume  -0.006             0.588**              0.796***
                        (0.272)            (0.292)              (0.209)

Post Volume            1.220***            0.564**              -0.389**
                        (0.235)            (0.263)              (0.193)

Incumbency                                 1.222***             -0.260
                                           (0.239)              (0.181)

Population                                  0.134               -0.069
                                           (0.098)              (0.073)

Google Search Trends                                            0.114***
                                                                (0.033)

Wikipedia Page Views                                            0.506***
                                                                (0.026)

Constant               -0.114              -2.038                0.112
                        (0.170)            (1.461)              (1.043)

-------------------------------------------------------------------------------
Observations             540                 540                  540
R2                      0.459               0.485                0.741
Adjusted R2             0.457               0.481                0.738
Residual Std. Error  2.114 (df = 537)   2.067 (df = 535)    1.468 (df = 533)
F Statistic       227.864*** (df = 2; 537) 125.808*** (df = 4; 535) 254.098*** (df = 6; 533)
===============================================================================
Note:                                              *p<0.1; **p<0.05; ***p<0.01
```

**Figure 6. Post Type Models: Campaign-Related Posts and Political Donations (Sum and Count)**

```
===============================================================================
                                         Dependent variable:
                        -------------------------------------------------------
                                         Donations (Log, 5000s)
                             Baseline      Partial Controls         Full
                               (1)              (2)                  (3)
-------------------------------------------------------------------------------
Issue-Related Volume         0.944***          0.745***            0.541***
                             (0.135)           (0.141)             (0.111)

Post Volume                  -0.059             0.013              -0.259***
                             (0.114)           (0.114)             (0.091)

Incumbency                                     0.604***            -0.185
                                               (0.143)             (0.120)

Population                                     0.123**             0.030
                                               (0.061)             (0.049)

Google Search Trends                                                0.044*
                                                                   (0.023)

Wikipedia Page Views                                                0.285***
                                                                   (0.017)

Constant                     0.050            -1.819**             -0.864
                             (0.106)           (0.910)             (0.714)

-------------------------------------------------------------------------------
Observations                  540               540                 540
R2                            0.457             0.476               0.680
Adjusted R2                   0.455             0.472               0.677
Residual Std. Error    1.298 (df = 537)    1.278 (df = 535)    1.000 (df = 533)
F Statistic            225.767*** (df = 2; 537) 121.401*** (df = 4; 535) 189.159*** (df = 6; 533)
===============================================================================
Note:                                                *p<0.1; **p<0.05; ***p<0.01

===============================================================================
                                         Dependent variable:
                        -------------------------------------------------------
                                         Donations (Count)
                             Baseline      Partial Controls         Full
                               (1)              (2)                  (3)
-------------------------------------------------------------------------------
Issue-Related Volume         1.128***          0.888***            0.507***
                             (0.215)           (0.225)             (0.164)

Post Volume                  0.307*            0.401**             -0.063
                             (0.181)           (0.183)             (0.135)

Incumbency                                     0.735***            -0.657***
                                               (0.228)             (0.176)

Population                                     0.102               -0.082
                                               (0.098)             (0.073)

Google Search Trends                                                0.114***
                                                                   (0.033)

Wikipedia Page Views                                                0.494***
                                                                   (0.026)

Constant                     0.129            -1.444               0.302
                             (0.168)           (1.456)             (1.054)

-------------------------------------------------------------------------------
Observations                  540               540                 540
R2                            0.486             0.495               0.739
Adjusted R2                   0.484             0.492               0.736
Residual Std. Error    2.062 (df = 537)    2.045 (df = 535)    1.475 (df = 533)
F Statistic            253.384*** (df = 2; 537) 131.348*** (df = 4; 535) 250.996*** (df = 6; 533)
===============================================================================
Note:                                                *p<0.1; **p<0.05; ***p<0.01
```

**Figure 7. Post Type Models: Issue-Related Posts and Political Donations (Sum and Count)**

Variation in the relative effects of specific types of content based on alternative specifications of donation receipts suggest future work should continue to distinguish sums from counts when assessing electorally-related payoffs of social and other digital media, and on a more practical level, indicate campaigns may find different types of content of greater utility based on the sought-after payout. Through these insights, based on methodological innovation in an area with a shortage of empirical work, this work paves the way for continued investigation into how an Internet-mediated trend in political campaigning is shaping and can be used to shape a longstanding ingredient of electoral success.

One should note however that Twitter and Facebook users are a minority of society (Blank, Graham, & Calvino, 2017) and often not of the demographic to actively engage in electoral behavior. While this would not negate the findings, and specific mechanisms for the discovered relationship are not examined empirically herein, it might raise the possibility that something other than user reception to candidate content on social media explains the relationship. However, multiple empirical looks have found that Twitter users are in fact the 'ideal subpopulation' with whom elites might desire to communicate, given they are very likely to turn out to the polls, interested in politics, and wealthy enough to contribute to campaigns (Bode & Dalrymple, 2016).

## Acknowledgements
We thank Laura Curlin for her assistance in gathering and cleaning the donation dataset and Jonathan Bright for insightful discussions.

## Funding Statement
TY was partially supported by The Alan Turing Institute under the EPSRC grant EP/N510129/1.

## Declaration of Conflicting Interest
The authors declare no potential conflicts of interest with respect to the research, authorship, and/or publication of this article.

## Data and Replication Information Meta-Document
The code used to access and analyze all data involved in this research may be found here: https://figshare.com/s/ff85b0a22fe470de5e80.

## Supplementary Information
See below.

# Supplementary Information for
## Social Media, Money, and Politics:
## Campaign Finance in the 2016 US Congressional Cycle

Lily McElwee[1] and Taha Yasseri[1,2]

[1]Oxford Internet Institute, University of Oxford, Oxford, UK
[2]Alan Turing Institute, London, UK

## Table of Contents

# **DATA**

## *Collection*

We collect data on the number and sum of donations received and the social media activity conducted by general election senatorial candidates during the six-week period prior to Election Day in the 2016 cycle. We pair social media data in a given period with donation data one period later to account for the temporal element in reactive donations and get at causation rather than simple correlation. Hence, the ultimate dataset for regression analysis consists of five pairs of periods, with social media data directly before the election not considered. Due to proximity with the studied time, there is no central repository of such data, so accessing and classifying form a major component of this study.

## Donations

The Federal Election Commission (FEC) provides searchable donation records based on reports from registered campaign committees. While there are application programming interfaces (APIs) to access FEC data for a specific set of candidate campaign accounts, we collaborate with MapLight to retrieve the total sum and count of donations received by the candidates with campaign accounts in the five-week period prior to Election Day.[1] The customized dataset includes 83 of 108 candidates. Twenty-five candidates are missing donation information altogether across the five periods. Candidates are not required to file with the FEC if they receive or spend under $5000, so the limited totals of these candidates is the most likely explanation for absent records, especially since data collection falls sufficiently after reporting deadlines.

## Social Media

We use raw Facebook posts and tweets as a basis for the social media variables. We manually search within each social media platform for candidate handles and check for additional accounts via Google. Many candidates have more than one Twitter or Facebook account.[2] For each candidate, we include all accounts found on a given platform to capture the entirety of his or her Twitter or Facebook presence.

96 of 108 candidates have a Twitter or Facebook presence. Among these, 86 have at least one Twitter profile, and 78 have at least one Facebook profile. 35 have at least two Twitter profiles, and 28 have at least two Facebook profiles.

With this set of handles, we use the Facebook Graph API and the Tweepy API package[3] in Python to collect the id, time and date, and text of the post for both Facebook and Twitter for a given screen name. Combining the entire base of accessible posts results in 2,248 Facebook entries and 18,243 Twitter entries, totalling 20,491 posts.

## Controls

*Google Trends*. we collect daily data on Google search trends over the past year for each candidate's full name via the gtrendsR package in R.[4] The raw data range from 0 to 100, since search volume obtained through Google Trends represents the traffic for a specific keyword relative to all queries submitted in Google, normalized on a 0-100 scale (where 100 represents the peak of relative search volume obtained per keyword during a given period).

*Wikipedia*. We manually search for candidates' Wikipedia entries; of 108 candidates, 71 have dedicated Wikipedia articles. The list provides a basis for collection of daily data on Wikipedia page views for each available article, which we do in R through the wikipediaR package (Bar-Hen, Baschet, Jollois, & Riou, 2016).[5]

---

[1] Maplight: https://maplight.org/.
[2] As an example, Hawaii Senator Brian Schatz possesses personal ('brianschatz'), office ('SenBrianSchatz'), and campaign ('SchatzforHawaii') accounts.
[3] For Facebook API, see here: https://developers.facebook.com/docs/graph-api. For Tweepy, the Twitter API package utilized, see here: http://www.tweepy.org/.
[4] gTrendsR package: https://cran.r-project.org/web/packages/gtrendsR/gtrendsR.pdf.
[5] For wikipediaR package, see here: https://cran.r-project.org/web/packages/WikipediaR/WikipediaR.pdf.

*Incumbency*. We collect information on whether a candidate is an incumbent from Ballotpedia.[6]

*Population*. For each candidate, we collect data on the population of their state from 2016 census figures.

## *Filtering*

### Donations

To address the fact that some candidates are missing donation records due to total sums below $5000, we put donation data for all candidates on a $5000-scale. In other words, if a candidate raised $10,000 in period three, the donation sum variable would be 2. This allows us to set the sum per period for the candidates without FEC records to zero, since the total was below $5000 in each period. This approach preserves as data points those individuals without an FEC record.

### Social Media

#### *Volume*

Raw tweets and Facebook posts serve as the basis for our first independent variable: social media volume per candidate-period. We aggregate the individual posts to get a count of how many times each candidate posted across platforms for each of the six periods prior to Election Day. For those without any sort of social media presence, we set post volume to zero in order to preserve the observations in the dataset. This is preferable to eliminating the observations altogether, and any model should retain predictive power for very low levels of social media activity as well as higher ones.

#### *Content*

The scraped posts also serve as the basis for the second genre of independent variable: type of post. To classify posts by topic, we use latent Dirichlet allocation (LDA) topic modeling, which postulates a latent structure in a corpus and represents each document as a distribution over topics (Steyvers and Griffiths, 2007). The advantages of this unsupervised method over alternative classification approaches for political content are numerous. Because it does not require prior labeling or annotation of text, topic modeling detects topics rather than begin with certain ones in mind, thereby avoiding oversight of ones that may be present in the corpus. The automatic nature of topic modeling lends it scale: the approach is better able to deal with higher volumes of content than hand coding, and such volume increasingly characterizes studies of social media (Blei, 2012).

On a practical level, classifying content via topic modeling involves cleaning the corpus of social media messages, determining the optimal number of topics, running the model, setting a threshold by which to classify a document as a certain topic, and developing a classification scheme by which to group topics thematically.

*Corpus cleaning*. Cleaning the data in preparation for topic modeling is an extensive process. After converting all the words to lowercase, removing both punctuation and numbers, and stripping whitespace, we conduct steps that typically come prior to processing tokenized texts: stemming and stopword removal. The former trims all terms down to their morphological roots (changing "november" to "novemb", for instance) and the latter eliminates meaningless but frequent terms such as "and", "or" and "but". We then adopt a tailored, iterative process involving text checks; this requires removing domain-related terms, hex terms, nonwords, custom stop words, name and state hashtags, candidate names, state names, candidate handles, and the top five most frequent terms (due to a cutoff at this value in term frequency), and terms that appear just once in the corpus. This process is summarized in Figure SI.1.

---

[6] 2016 US Senate Elections, Ballotpedia: https://ballotpedia.org/United_States_Senate_elections,_2016.

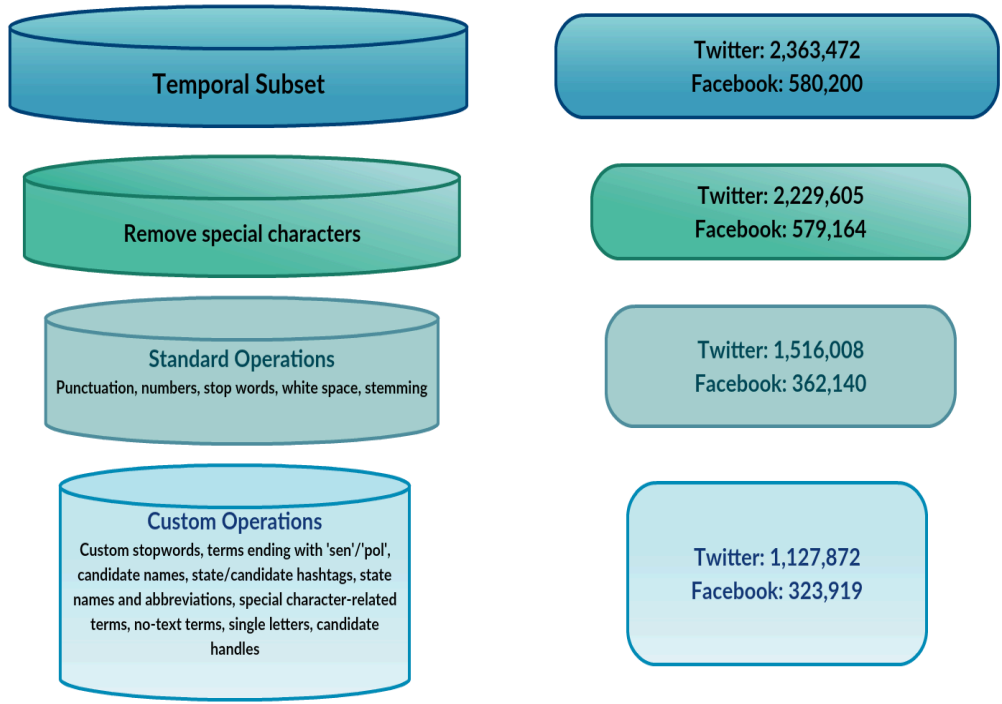

**Figure SI.1. Corpus Cleaning Operations**. *Total character count in the Twitter and Facebook posts following each step in the cleaning process.*

## Controls

*Google Search Trends*: To retrieve absolute rather than relative numbers, we scale Google search trends using average monthly search volume data from Adwords.[7] We create a ratio of average monthly search volume via Adwords to that via Google Search Trends by dividing the former value by the calculated monthly average Trends value based on a year's worth of data. For instance, Arn Menconi shows an average monthly Trends number of 56 over the past year and an average search volume of 1600 via Adwords, so the relevant ratio would be 29. We then scale each daily view in the relevant week-long periods based on this candidate-specific value. One caveat is that the scale is based on international views while we are specifically interested in US views, but the approach is reasonable given that most views for US senatorial candidates are likely to be from the US.

*Wikipedia Page Views*: We aggregate the daily data by the relevant periods in order to have Wikipedia page view data for each of the candidate-period observations for which it is available. For those without a Wikipedia article, we set the number of views to zero for each period.

*Incumbency and State Population.* The datasets for these controls come pre-formatted and are annual, so do not require any subsetting. We assume that population remains roughly similar across weeks leading up to Election Day.

## *Missing Data*

### Donations

Twenty-five candidates are missing donation information altogether across the five periods for which this data is collected. These include the following candidates: Allen Buckley, Bill Barron, Bill Bledsoe, Brian Chabot, Cris Ericson, Edward Clifford, Eric Navickas, Frank Gilbert, Gary Swing, Jeff Russell, Jerry Trudell, Jim Lindsay, John Heiderschet, Jonathan Dine, Kent McMillen, Mike Workman, Pete Diamondstone, Phil Anderson, Richard Lion, Robert Marquette, Robert Garrard, Robert Murphy, Scott Summers, Stoney Fonua, Tom Sawyer. Those with paper filings but no reported itemized contributions during the periods selected include: Bill Bledsoe, Bill Barron, and Phil Anderson. Many of the remaining candidates have an FEC record but no financial data for the 2015-16 cycle or other elections.

---

[7] We are able to access this information by setting up an Adwords account and searching for each candidate's name here: https://adwords.google.com/um/GetStarted/Home?__u=1110451079&__c=6445738689&authuser=0#oc.

## Social Media

In some cases, there are candidates for whom some social media activity occurred, but either Facebook or Twitter information is not retrieved. In several cases, such as Joe Heck and Stoney Fonua for Facebook, the candidate appears to have once had a social media presence that is no longer available, due to account deletion and archiving. In other cases, temporal subsetting does not include the available posts, such as with the SenatorCortezMasto and SenEvanBayh handles for Facebook and the ArnMenconi and ShelbyForSenate handles for Twitter. Often, these candidates are included in the analysis because there is information about posting through still-accessible parts of these candidates' social media presences, but it is worth noting that this presence may in fact have been larger than is represented in the calculated social media activity figures. A full list of eliminated handles is provided here.

Profiles for which no Facebook data is accessible:
- Profiles (rather than pages):
    - 'rscrumpton'
    - 'ray.metcalfe.16'
    - 'arnmenconi'
    - 'robinlaverne'
    - 'michael.k.workman'
    - 'eric.navickas'
    - 'edward.clifford.56'
    - 'jerry.trudell.7'
- Archived/Not Available
    - 'wiesnersenate'
    - 'heck4nevada'
    - 'CallahanForOregon'
    - 'senatortoomey'
    - 'stonefonua'
    - '2016crisericson'
- Business Page
    - 'Mike-Crapo-For-US-Senate-286049384763373'
    - 'Tom-Jones-for-US-Senate-1752874281645766'

Profiles for which no Twitter data is available:
- No tweets
    - tonyg4senate
    - Callahan4Oregon
- No posts in time frame
    - ArnMenconi
    - DickBlumenthal
    - SenEvanBayh
    - ShelbyForSenate

## *Candidate Information & Posts*

Full dataset is listed as "2016 Senatorial Candidates: Candidate Information & Posts" in the following repository: https://figshare.com/s/4183e5df4e1b959701a5. Figure SI.2 provides a summary of included variables.

| Category | Variable | Type | Data |
|---|---|---|---|
| Information | name | Categorical | Nominal |
| Temporal | period | Numerical | Discrete |
| Information | abb | Categorical | Nominal |
| Information | state | Categorical | Nominal |
| Social-Media Related | text | Categorical | Nominal |
| Social-Media Related | date | Numerical | Discrete |
| Social-Media Related | docid | Categorical | Nominal |
| Social-Media Related | platform | Categorical | Nominal |
| Social-Media Related | handle_type | Categorical | Nominal |
| Control-Related | Wikipedia. | Categorical | Nominal |
| Information | party | Categorical | Nominal |
| Control-Related | incumbency | Categorical | Ordinal |
| Control-Related | population | Numerical | Continuous |
| Links in the post | out_links | Categorical | Nominal |
| Links in the post | contains_link | Categorical | Ordinal |
| Social Media-Related | maxname | Categorical | Ordinal |
| Social Media-Related | morethanone | Categorical | Ordinal |
| Social Media-Related | empty | Categorical | Ordinal |

**Figure SI.2. Variables in the Candidate Information & Posts Dataset.**

## *Regression*

Full dataset is listed as "2016 Senatorial Candidates: Social Media Volume/Type & Campaign Donations" in the following repository: https://figshare.com/s/e271230e5a52e7c1d57e. Figure SI.3 provides a summary of included variables.

| Category | Variable | Type | Data |
|---|---|---|---|
| Information | name | Categorical | Nominal |
| Temporal | period | Numerical | Discrete |
| Information | abb | Categorical | Nominal |
| Information | state | Categorical | Nominal |
| Control | wiki_views | Numerical | Continuous |
| Information | FECID | Categorical | Nominal |
| Donation-Related | cash_total | Numerical | Continuous |
| Donation-Related | cash_count | Numerical | Continuous |
| Donation-Related | scaled_total | Numerical | Continuous |
| Social Media-Related | post_volume | Numerical | Continuous |
| Social Media-Related | none | Numerical | Continuous |
| Social Media-Related | topic1 | Numerical | Continuous |
| Social Media-Related | topic2 | Numerical | Continuous |
| Social Media-Related | topic3 | Numerical | Continuous |
| Social Media-Related | topic4 | Numerical | Continuous |
| Social Media-Related | topic5 | Numerical | Continuous |
| Social Media-Related | topic6 | Numerical | Continuous |
| Social Media-Related | topic7 | Numerical | Continuous |
| Social Media-Related | topic8 | Numerical | Continuous |
| Social Media-Related | topic9 | Numerical | Continuous |
| Social Media-Related | topic10 | Numerical | Continuous |
| Social Media-Related | maxname | Categorical | Nominal |
| Social Media-Related | issue_related | Numerical | Continuous |
| Social Media-Related | campaign_related | Numerical | Continuous |
| Social Media-Related | issue_share | Numerical | Continuous |
| Social Media-Related | campaign_share | Numerical | Continuous |
| Information | party | Categorical | Nominal |
| Control-Related | Wikipedia. | Categorical | Nominal |
| Control | incumbency | Categorical | Ordinal |
| Information | outcome | Categorical | Ordinal |
| Control | population | Numerical | Continuous |
| Control | scaled_gst | Numerical | Continuous |
| Donation-Related | donations_sum_lag | Numerical | Continuous |
| Donation-Related | donations_count_lag | Numerical | Continuous |
| Dependent | log_sum_lag | Numerical | Continuous |
| Dependent | log_count_lag | Numerical | Continuous |
| Independent | log_post_volume | Numerical | Continuous |
| Independent | log_campaign | Numerical | Continuous |
| Independent | log_issue | Numerical | Continuous |
| Control | log_gst | Numerical | Continuous |
| Control | log_wp | Numerical | Continuous |
| Control | log_population | Numerical | Continuous |

**Figure SI.3. Variables in the Regression Dataset.**

# TOPIC MODELING

*Corpus Cleaning Operations*

Here we describe in greater detail each cleaning step taken to prepare the corpus for topic modeling.

**Standard Operations**
- **Lowercase.** Convert all terms to lowercase.
- **Punctuation.** Remove standard punctuation terms.
- **Numbers.** Remove numbers.
- **Stopwords.** Remove standard English stop words.[8]

---

[8] The full list of stopwords used in the R operation is provided here: http://jmlr.csail.mit.edu/papers/volume5/lewis04a/a11-smart-stop-list/english.stop.

- **Whitespace.** Convert multiple whitespace characters into a single blank (Feinerer, 2017).
- **Stemming.** Apply Porter's stemming algorithm to stem words.[9]

**Custom Operations**
- **Domain words.** Many of the posts include links. Because they are largely shortened Twitter links (e.g. 'm2umndgs86'), they are not informative regarding the purpose of the post, and therefore demand removal. We remove these by first identifying all the end components of the Twitter links, which involves extracting all links from the each document using regex and removing all non-alphanumeric characters and numbers. Once we create a vector of all the domain ends, removing these full links is simple, since the operation to remove special characters above splits each full link into several components and it is simple to include http:// and https:// and "t" and "co" as nonword terms, such that they are removed when that command is run.
- **Hex terms.** These are used to indicate a backslash or other special characters, but show up as xef or xbn. We remove these because they are meaningless.
- **Nonwords.** Any words observed in manual examination that do not add meaningful content, such as 'schriok' (a name), are removed. As noted above, this includes collections of letters that are created by the removal of numbers, links, and punctuation in the initial cleaning (such as "https" and "www" and 'ly').
- **Custom stop words.** These are words that do not have meaning but are also not proper terms such as names. They include terms such as "also", "other", "another", "ill", "im", "ive". Some of them are created by the removal of punctuation in the first cleaning step.
- **Candidate names (first and last).** We remove the candidate's first and last names, because they do not add meaningful information in the quest to identify patterns.
- **Name/State Hashtags.** Many candidates use hashtags to identify the race in which they are running, such as "#AZSen" by Democratic senatorial challenger in Arizona Ann Kirkpatrick and "laelex" by Republican challenger John Neely Kennedy in Louisiana. These terms do not offer meaningful information, since they vary by state/candidate.
- **State names.** We remove the names of specific states, such as "California", because these do not help in identification of patterns.
- **Candidate handles.** We remove the Twitter and Facebook handles of the candidates. We keep all other handles, such as realdonaldtrump, hillaryclinton, and drjillstein, since appearance of these terms offers meaningful information – such as, for instance, whether the candidates are responding to, supporting, or discussing federal candidates.
- **Letters.** We remove all individual letters of the alphabet.
- **Most frequent terms.** We remove the five most common topics in the document, based on a cutoff in frequency when all terms are examined. These include: "vote", "senat", "thank", "today", "work." Because these terms appear over and over again in the text, they do not improve the chances of distinguishing messages based on differential content.
- **Least frequent terms.** The least frequent terms do not aid identification of patterns in the documents, so we remove all terms that appear just once in the corpus. This involves eliminating 10,183 terms, which encompasses 53.19% of the total terms.

  On a technical level, removal of the 306 entries that are empty following cleaning means that the output of the topic modeling tool does not align directly in terms of document numbers with the posts listed in the original file. While we could simply remove the posts that display no text or become empty in the cleaning process from the original file, this would lose valuable information about the volume of activity by a candidate. Instead, after cleaning the data, we record each document's original position in the full corpus. Alongside information about where the documents are in the non-empty corpus, retaining this information allows us to align the output of the topic modeling tool with the correct row positions in the candidate information file. Hence, the candidate file with classification information remains 20,491 entries but there are 306 cases in which no classification information whatsoever is provided, since these are empty entries either before or after cleaning.

---

[9] The paper originally presenting the algorithm is Porter (1980). Documentation for the text mining package to which the stemming algorithm belongs in R is here: https://cran.r-project.org/web/packages/tm/vignettes/tm.pdf.

## Topic Modeling Output - Words in Topics

| Topic | Top 10 Words |
|---|---|
| 1 | great call morn plan chang meet stop check obamacar win |
| 2 | offic student visit secur news learn educ home school open |
| 3 | elect join women volunt voter counti earli campaign sign ballot |
| 4 | trump polit ad presid clinton race republican donald sen deserv |
| 5 | famili state year veteran countri nation mani tax honor govern |
| 6 | american continu care read health tour stop afford lead end |
| 7 | work communiti issu import leader talk hard discuss togeth colleg |
| 8 | debat candid endors stand live tonight parti record night run |
| 9 | time peopl poll tomorrow share futur life event big don |
| 10 | fight watch job proud thing good protect fund worker back |

**Figure SI.4. Top 10 Words Associated with Each Topic.**

## Topic Modeling Description

Among the most important steps in identifying the content of a document is determining which topics are addressed by that document (Griffiths and Steyvers, 2004). A given document, or message, can deal with multiple topics, and the words appearing in it reflect the particular set of topics it addresses. Latent Direct Allocation, the approach used in this paper, is an unsupervised statistical method that views topics as probability distributions over words and treats documents as probabilistic combinations of these topics, in order to model the contributions of different topics to a given document (Griffiths and Steyvers, 2004). This specific version of topic modeling involves the user entering a collection of documents and specifying a number of topics; hence, a major part of the process is cleaning the data – notably, 50-80% of a typical data scientist's role has been shown to involve data preparation (Lohr, 2014) – and then determining the optimum number of topics. As mentioned earlier, topic modeling is not a common approach in online political content analysis, due perhaps to the greater accessibility of approaches such as keyword search and hand-coding.

We remove original and cleaned posts that have no text. Null values arise because some original posts have no text (perhaps because they involve just a picture instead), or because the cleaning process drastically shortens them. Empty posts amount to 306 of 20,491 documents. While these are removed for topic modeling, since they do not add valuable qualitative information, we include these posts as part of the overall post volume for a given candidate-period and record them as 'none' type because they do not belong to a specific topic, a practice that minimizes extra categories for analysis and has precedent in topic.

Altogether, cleaning operations reduce the total number of words from 507,222 in the uncleaned message content to 185,010 in the cleaned corpus. Across platforms, the original messages range from 3 to 6603 characters, with a mean of 144 median of 139. The cleaned messages range from zero (inclusive of the empty posts) to 3,396, with a mean of 61 and median of 54. The average Facebook message is shortened from 258 characters to 134 characters, while the average Twitter message is shortened from 130 characters to 52 characters. With a cleaned corpus, we identify topics in the corpus and classify each post accordingly.

*Topic Number Optimization.* To identify an optimum number of topics, we use a function that loops over different topic numbers, calculates the log-likelihood of the model at each number of topics, and plots the result. We chose to assess the log-likelihood between 2 and 20 topics, in order to keep the number manageable for analysis, although there is no best practice with regard to the proportion of topics to words. The function used to create Figure SI.5 estimates a LDA model using 2000 iterations of Gibbs sampling for the document term matrix with the following control parameters: a document term matrix (all non-empty documents in the corpus), an 'nstart' value specifying the number of random restarts used during optimization (five here), and a seed vector with the same number of entries as the number of random restarts (Forte et al, 2017).

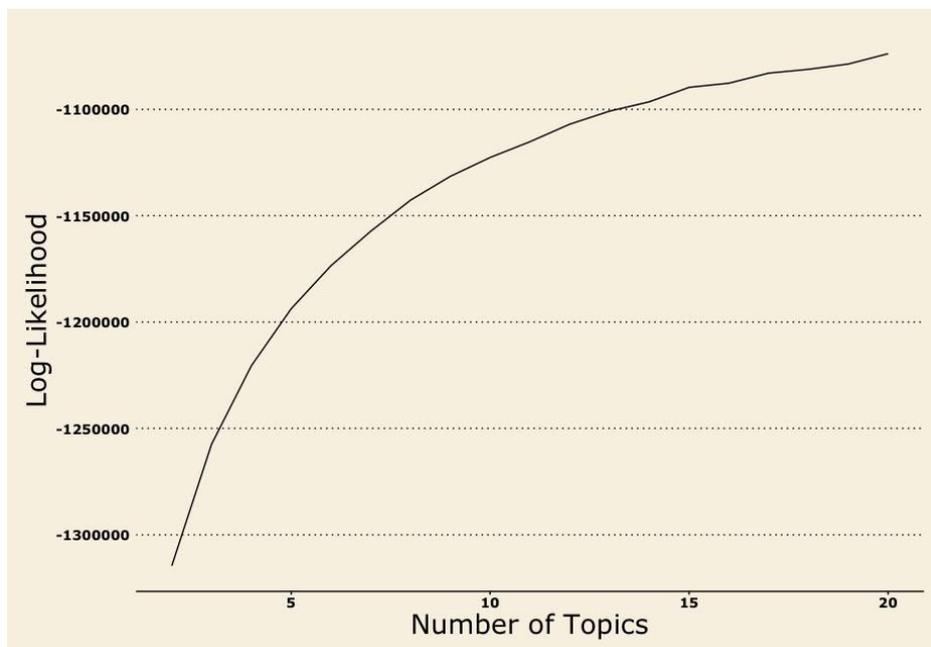

**Figure SI.5. Log-likelihood of data per topic number (2-20)**. *Optimal number of topics is 10, based on 'elbow' detection method.*

We use the 'elbow detection' method to identify the optimum number of topics, a technique with precedent in Twitter topic modeling (Steinskog, Therkelsen, & Gambaeck, 2017). As the marginal improvements in log-likelihood appear to decrease around 10 topics, we use this value as a cutoff. Having established 10 as the optimum number of topics, we specify this as a parameter in the graphical user interface tool and run the model.[10]

*Classifying Documents.* The process results in ten collections of words. Consistent with precedent (Messing, Franco, Wilkins, Cable, and Warshauer, 2014), topic labels are selected according to the words appearing most frequently in the posts classified by a specific topic. Figure SI.6 offers a summary.

| Topic | Name & Description |
|---|---|
| 1 | Campaigning metatalk, related to meetings, calls, and stops. |
| 2 | Policy issues, mostly related to education ('school', 'student', 'educ', 'learn') and security ('secur'). |
| 3 | Campaign topics, related to voting/election. |
| 4 | Campaign topics, related to partisan talk and discussion of federal topics, e.g. Democrat/Republican ('donald' and 'clinton'). |
| 5 | Policy issues, most related to veterans ('veteran', 'famili') and taxation ('tax') |
| 6 | Policy issues, mostly related to healthcare ('care', 'health', 'afford') but also others. |
| 7 | Policy issues, related to community ('leader', 'discuss', 'together', 'issu') and education ('colleg') |
| 8 | Campaign topics, such as endorsements and media appearances |
| 9 | Campaign topics, related to events ('time', 'event', 'tomorrow') and solicitation of volunteers. |
| 10 | Policy issues, mostly related to jobs ('job', 'worker', 'protect') |

**Figure SI.6. Topic Modeling Output: Qualitative Description**. *Brief description of the ten topics discovered through modeling.*

The algorithm presents each document as one topic or a share of many topics. The topic used to classify each document is selected as that with the highest description share for the document. Prior to classification of documents into topics, we define a threshold above which to consider classifications by plotting the distribution of all shares and of highest shares. Figure SI.7, of the highest shares, shows a first peak and natural cutoff at roughly 0.25, below which any shares can be inferred as noise. This is corroborated through examination of all shares (Figure SI.7), in which there is a peak at 0.24-0.25. Hence, we select 0.25 as the threshold for classifying documents according to the topic capturing the highest share. We set all other values (those below 0.25) to zero. While there are 106,800 nonzero topic shares in the original output of the topic modeling algorithm, this falls to 52,747 when the threshold of 0.25 is applied. The threshold reduces the total number of classified documents by 1766, such that there are this many with maximum shares that fall under the value (0.25) and hence cannot be classified according to any of the topics. Similar to posts with no text after the cleaning process, these are classified typologically in the 'none' category.

---

[10] The topic modeling tool utilized may be found here: https://code.google.com/archive/p/topic-modeling-tool/.

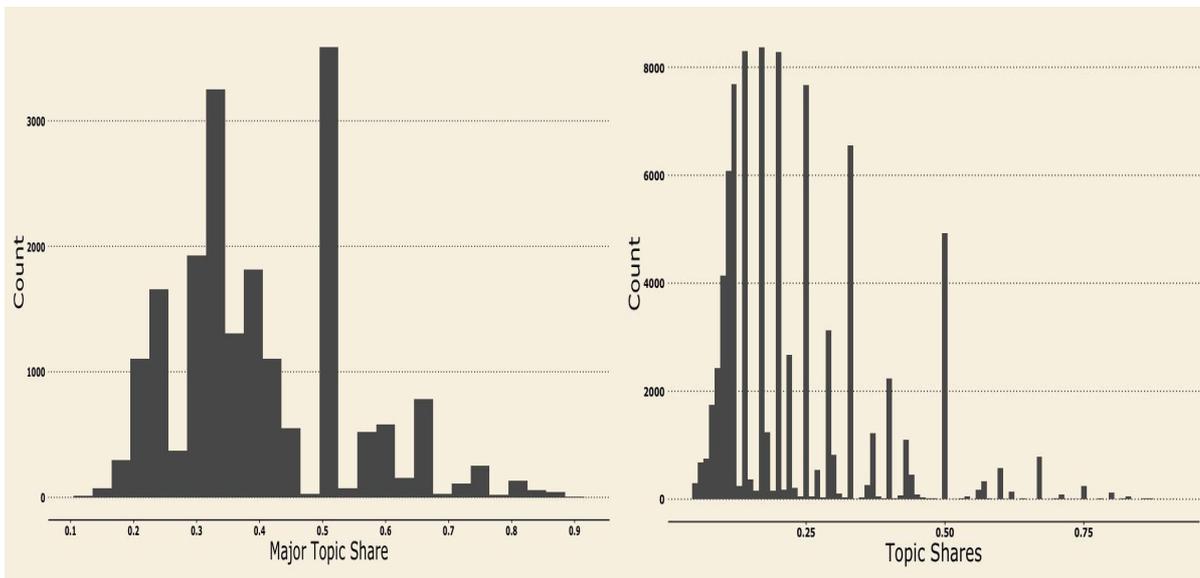

**Figure SI.7. Distribution of Shares Captured by Maximum Topic (Left) and Each Topic (Right)**. *Each document is represented as a probabilistic combination of topics. All shares are provided on the right, showing a natural cutoff at 0.24-0.25. Plotting the highest shares characterizing each document (left) in the corpus confirms a natural cutoff at 0.25.*

*Grouping Topics.* Analysing similarity between topics provides a basis for grouping individual topics and deriving a meaningful hypothesis about the relationship between fundraising and specific types of content. We use cosine similarity to identify the similarity of the automatically-generated topics in the LDA model, an approach with precedent in thematic topic aggregation (Messing et al, 2014). The metric runs between -1 and 1, where smaller angles (more similar topics) will be closer to 1 and larger angles (dissimilar topics) will produce a cosine similarity value nearer to 0; opposite topics are represented by a value of -1. Hence, a higher cosine similarity indicates closer connection between topics.

Figure SI.8 shows the distribution of cosine similarities across topics. Ignoring connections between the same topic, the range is 0.18 to 0.28 with a mean and median of 0.24.

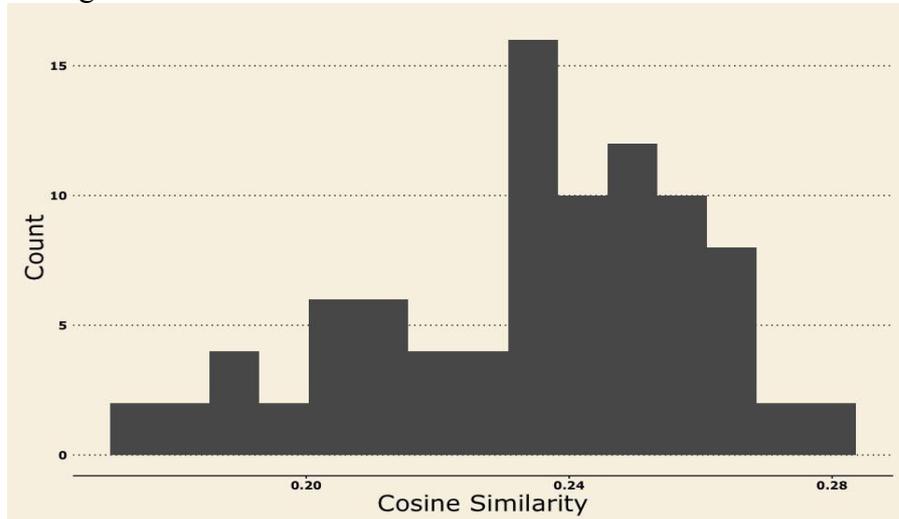

**Figure SI.8. Distribution of Cosine Similarity**. *Dramatic cutoff at 0.23, below which the cosine similarity values (or values measuring similarity between two topics) can be inferred as noise.*

The histogram (Figure SI.8) shows a rather dramatic cutoff at 0.23, below which the cosine similarity values can be inferred as noise. We convert all weightings below this threshold to zero in order to identify meaningful variation at the higher end of the spectrum. This leaves 30 unique pairs (of the original 45) for which the cosine similarity is above 0.23.

Setting the weights for all other pairs to zero, we visualize the topic similarity for those pairs above the cutoff in graphing platform Gephi.[11] The full network is composed of 10 nodes and 30 edges. Analyzing modularity via the Louvain community detection algorithm (Blondel, Guillaume, Lambiotte, and Lefebvre,

---

[11] Gephi Documentation: https://gephi.org/.

2008), we are able to detect communities of topics that are more similar to each other than to the others. Modularity is maximized with a resolution of 1.0; at this value, it is 0.07 whereas at 0.5 it is 0.06 (and three communities are detected) and anywhere upwards of 1.5 it is 0 (with 1 community detected). At resolution 1.0, there are two communities detected, as indicated by the color distinction in Figure SI.9.

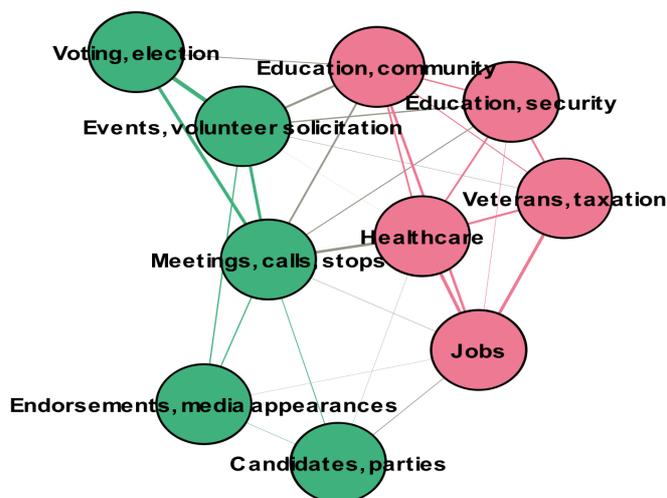

**Figure SI.9. Topic Network Graph**. *Based on cosine similarity weightings, the Louvain community detection method discovers two communities of topics, which qualitative examination shows to group thematically into campaigning-related (green) and issue-related (pink) posts.*

Visually, it is evident there are two main clusters of topics. Topics 1, 3, 4, 8, 9 form a community while topics 2, 5, 6, 7, 10 form a separate community. The former group all have campaigning-related elements while the other topics relate to policy issues, so the distinction enables us to test the broader hypothesis about the effect certain types of social media content have on fundraising success.

In order to form a variable for regression analysis, we calculate the number of a given candidate's posts per period that are classifiable as campaign- and issue-related. For each candidate-period, this involves summing the posts in topics 1, 3, 4, 8, and 9 for campaign-related posts and the posts in topics 2, 5, 6, 7, and 10 for the issue-related posts.

# DESCRIPTIVE STATISTICS

*Overall Distributions*

Distributions: Donations Sum and Counts

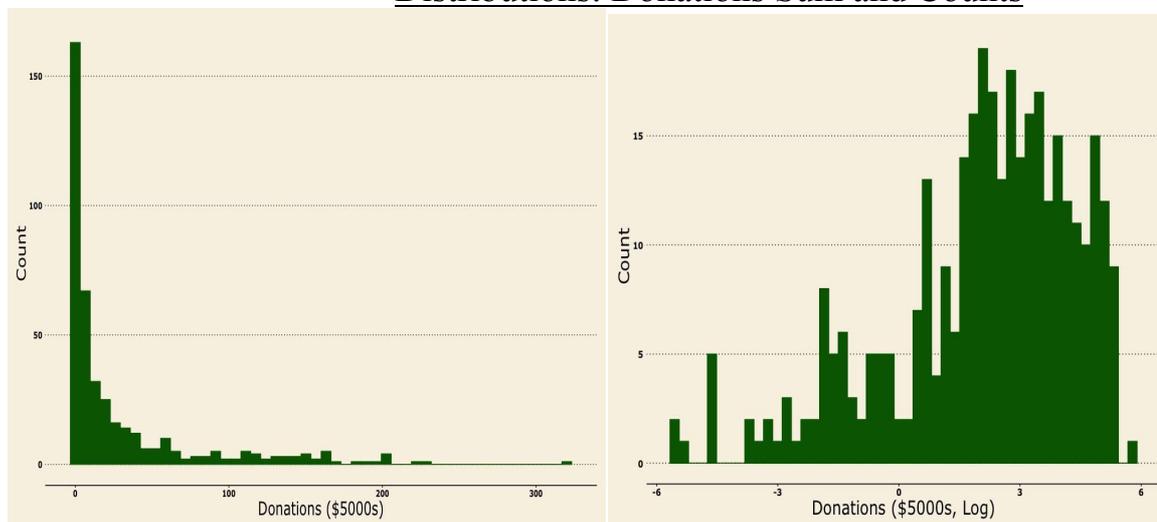

**Figures SI.10. Per-Period Donation Sums to US 2016 Senatorial Candidates ($5000s): Linear (Left) and Log (Right)**. *The distribution of donation sums per candidate-period is right-skewed and lognormal.*

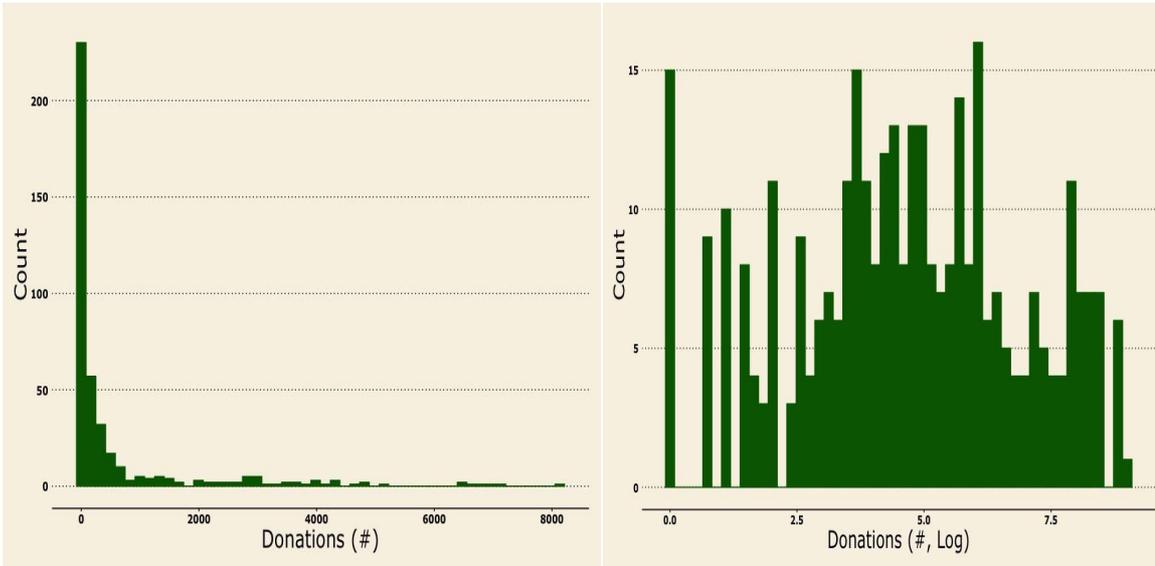
**Figures SI.11. Per-Period Donation Counts to US 2016 Senatorial Candidates (#): Linear (Left) and Log (Right)**. *The distribution of donation counts per candidate-period is right-skewed and lognormal.*

## Distributions: Twitter and Facebook Posts

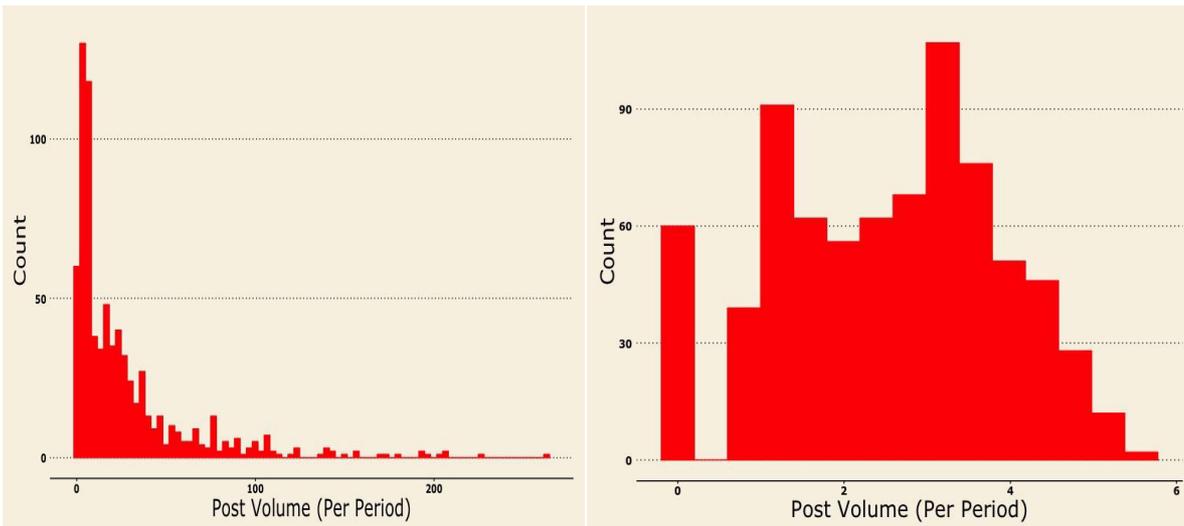
**Figures SI.12. Weekly Posts (Twitter + Facebook) by US 2016 Senate Candidates: Linear (Left) and Log-Transformed (Right)**. *The distribution of posts across all platforms per candidate-period is right-skewed and log-normal.*

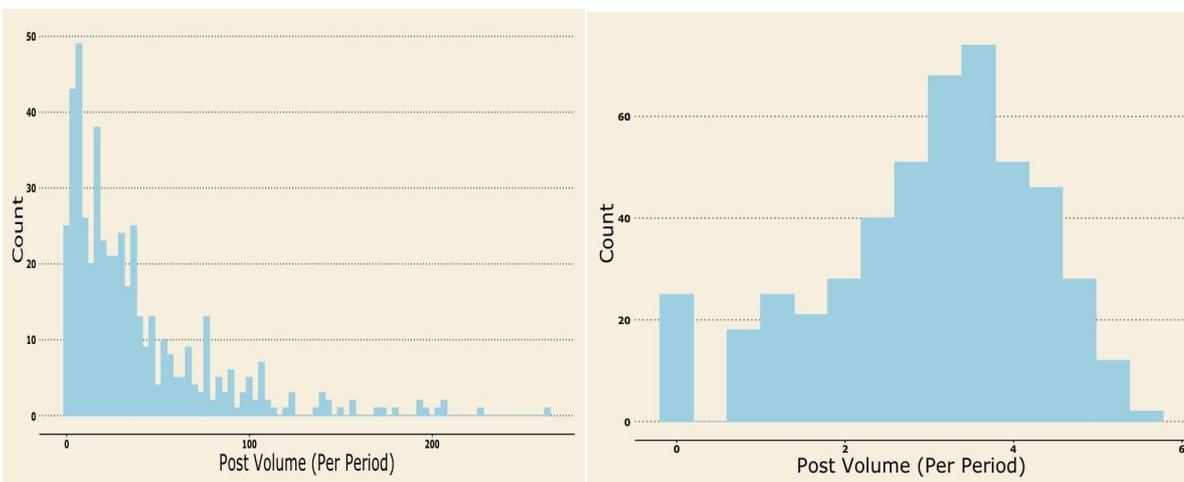
**Figures SI.13. Weekly Posts (Twitter) by US 2016 Senate Candidates: Linear (Left) and Log-Transformed (Right)**. *The distribution of tweets per candidate-period is right-skewed and log-normal.*

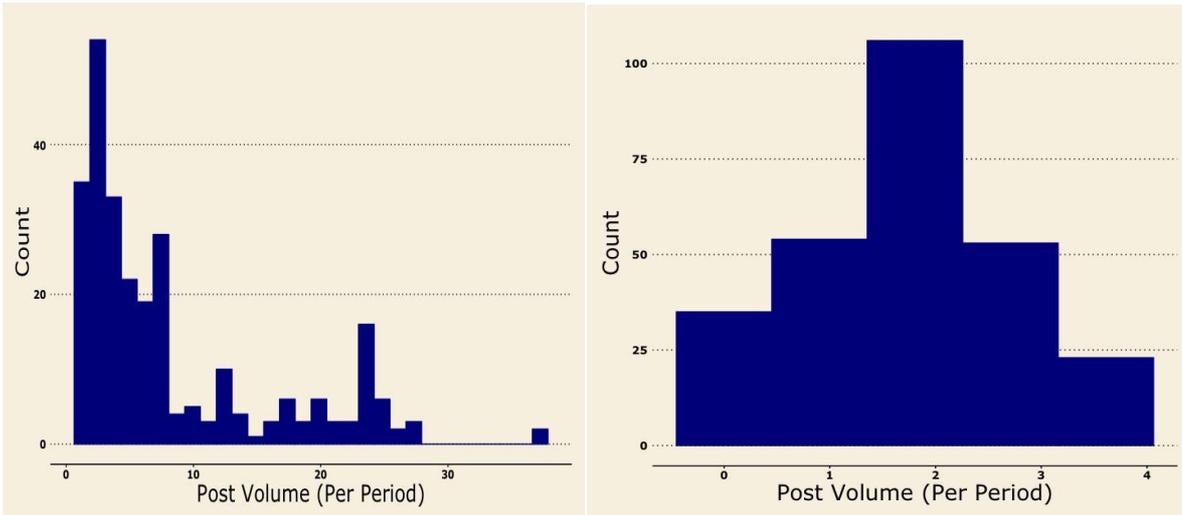
**Figures SI.14. Weekly Posts (Facebook) by US 2016 Senate Candidates: Linear (Left) and Log-Transformed (Right)**. *The distribution of Facebook posts per candidate-period is right-skewed and log-normal.*

## Distributions: Social Media Posts per Type

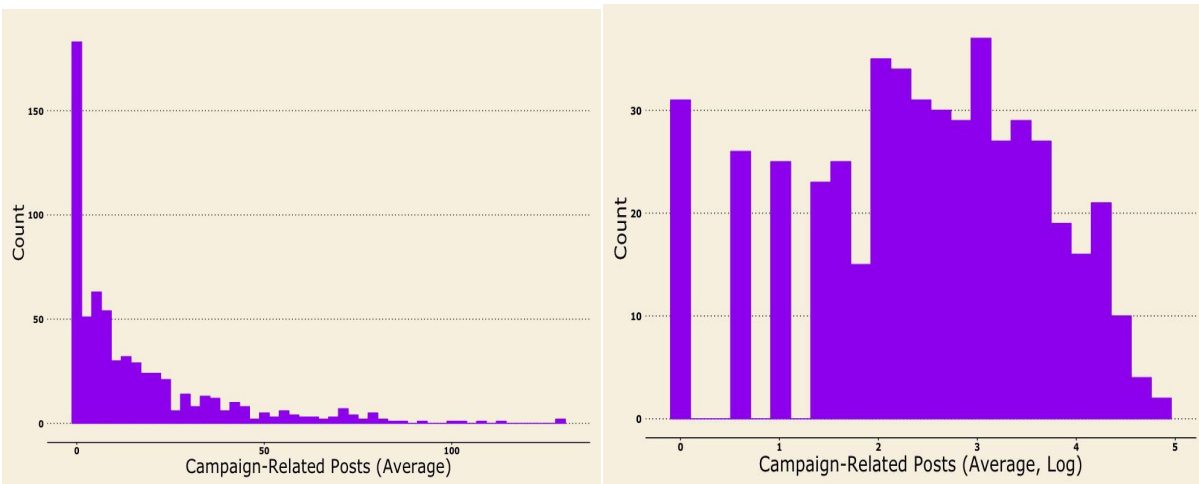
**Figures SI.15. Average Campaign-Related Posts Per Period: Linear (Left) and Log-Transformed (Right)**. *The average number of campaign-related posts per candidate across the six periods is lognormal.*

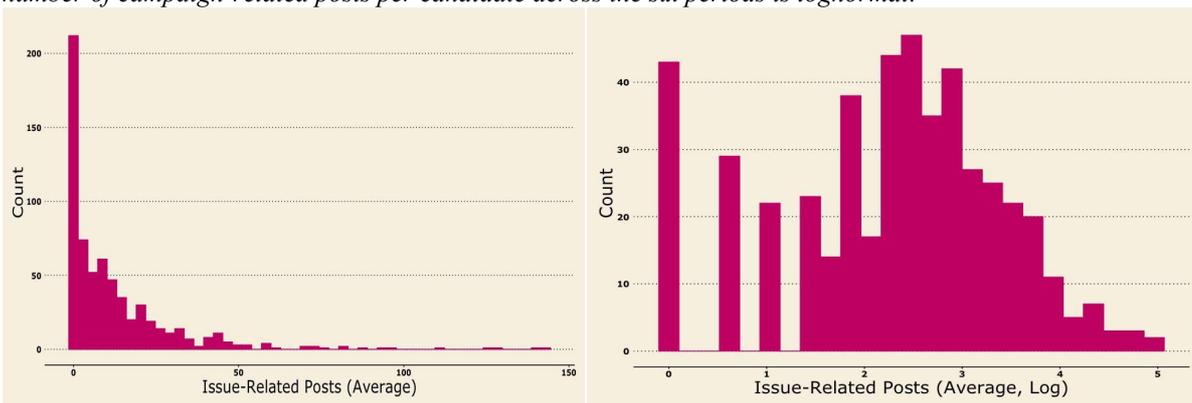
**FiguresSI.16. Average Issue-Related Posts Per Period: Linear (Left) and Log-Transformed (Right)**. *The average number of issue-related posts per candidate across the six periods is lognormal.*

## Distributions: Google Search

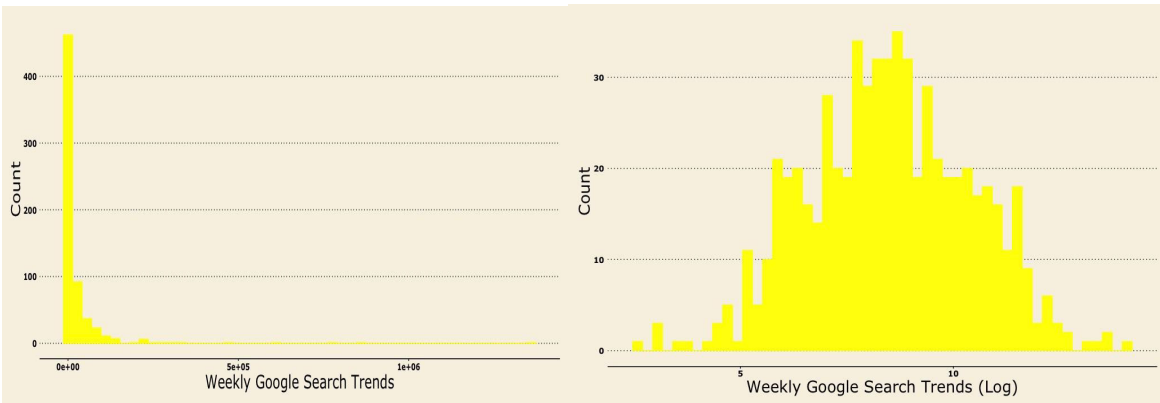

**Figures SI.17. Weekly Google Search Trends Per Candidate-Period: Linear and Log-Transformed**. *The distribution of Google search trends per candidate-period is right-skewed and log-normal.*

## Distributions: Wikipedia Page Views

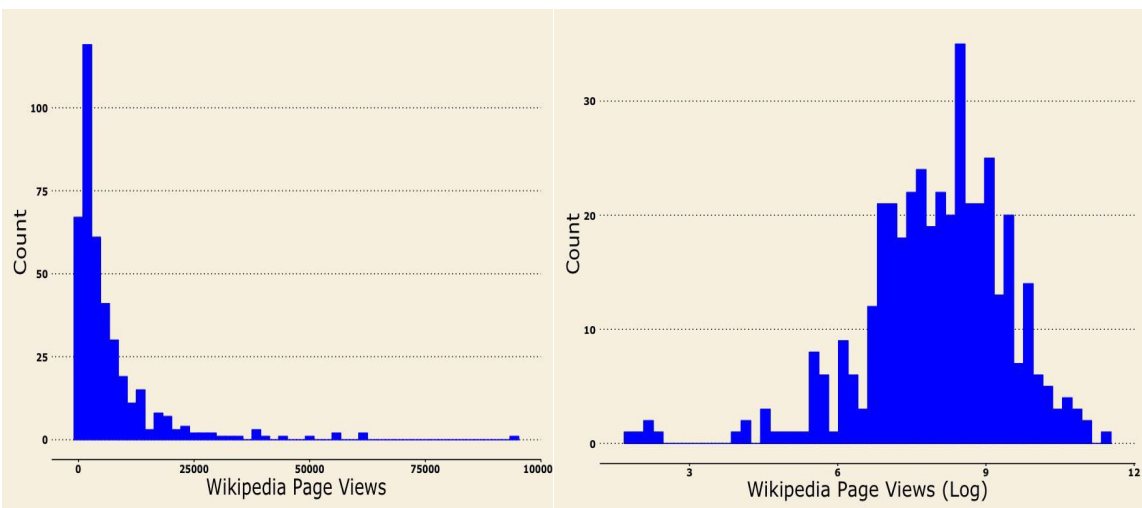

**Figures SI.18. Wikipedia Page Views, Per Candidate-Period: Linear and Log-Transformed**. *The distribution of Wikipedia page views per candidate-period is right-skewed and log-normal.*

## *Per-Period (week) Distributions*
### Google Search Trends, Wikipedia Page Views

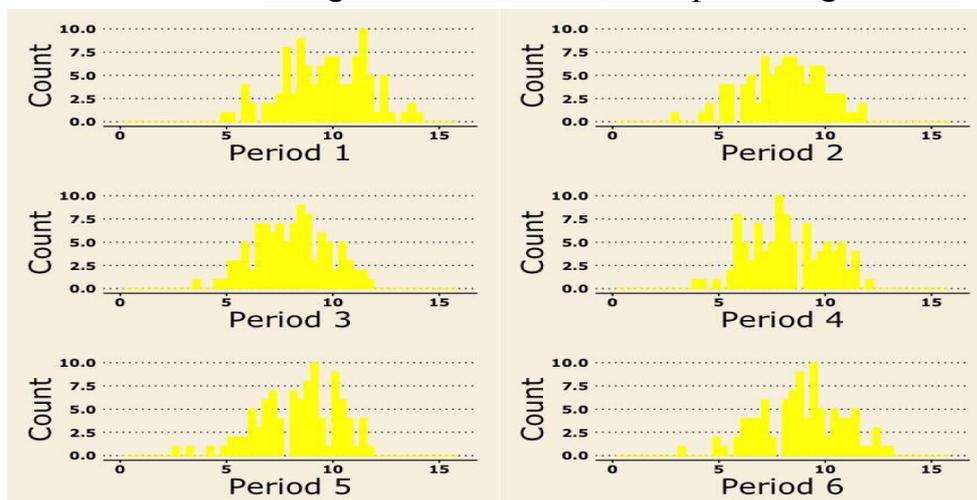

**Figure SI.19. Log-Transformed Distributions of Weekly Google Search Trends Across All Candidates, by Period.**

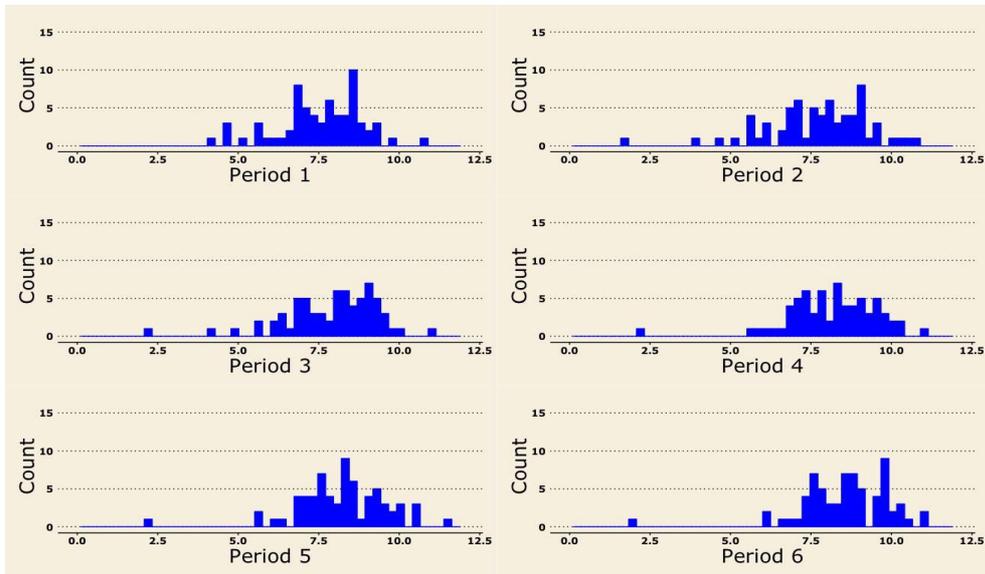

**Figure SI.20. Log-Transformed Distributions of Weekly Wikipedia Page Views Across All Candidates, by Period.**

## *Scatterplots*

Scatterplots reveal non-linear relationships when each independent variable is plotted against each dependent variable. Specifically, "residuals with a bow shape and increasing variability indicate that a log transformation...is required" (Wonnacott and Wonnacott, 1972, 463). The graphs below show the benefit of log transformation.

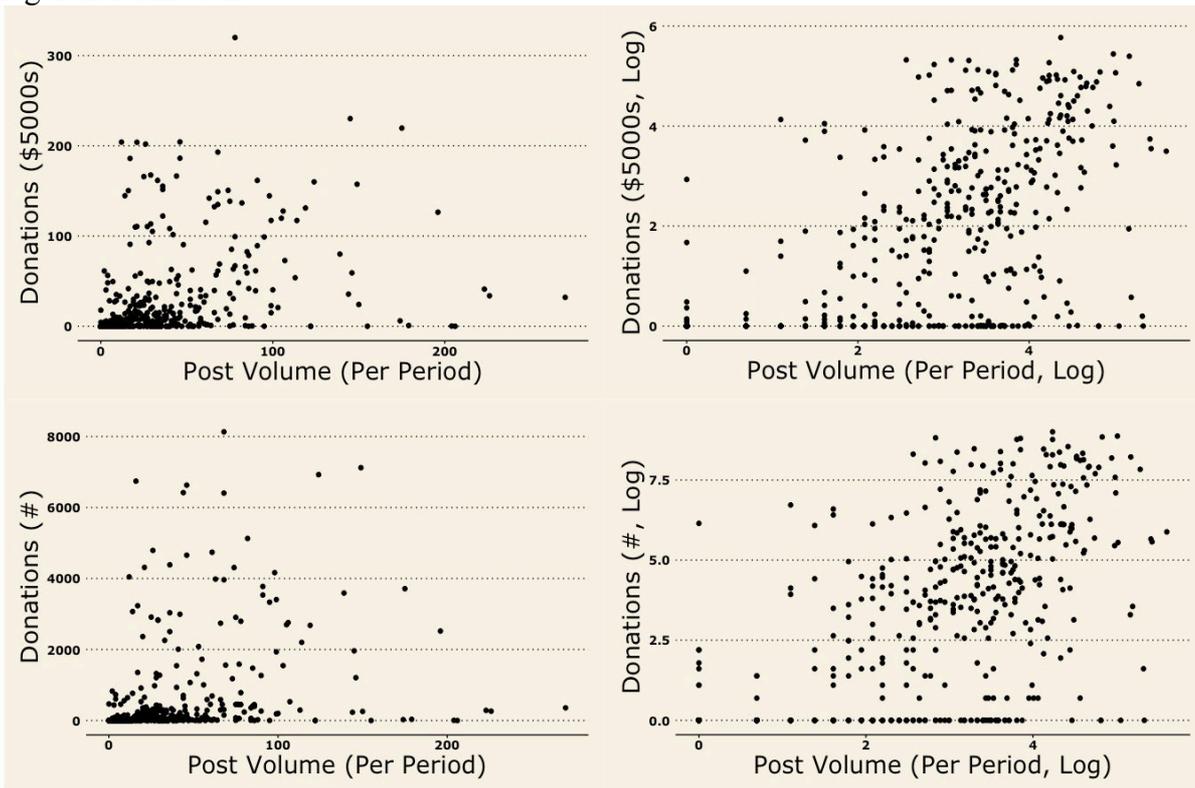

**Figure SI.21. Scatterplot of Post Volume versus Donations, Linear (Left) and Log-Transformed (Right).** *The relationship is not linear when in lin-lin form, suggesting the need for log-transformation.*

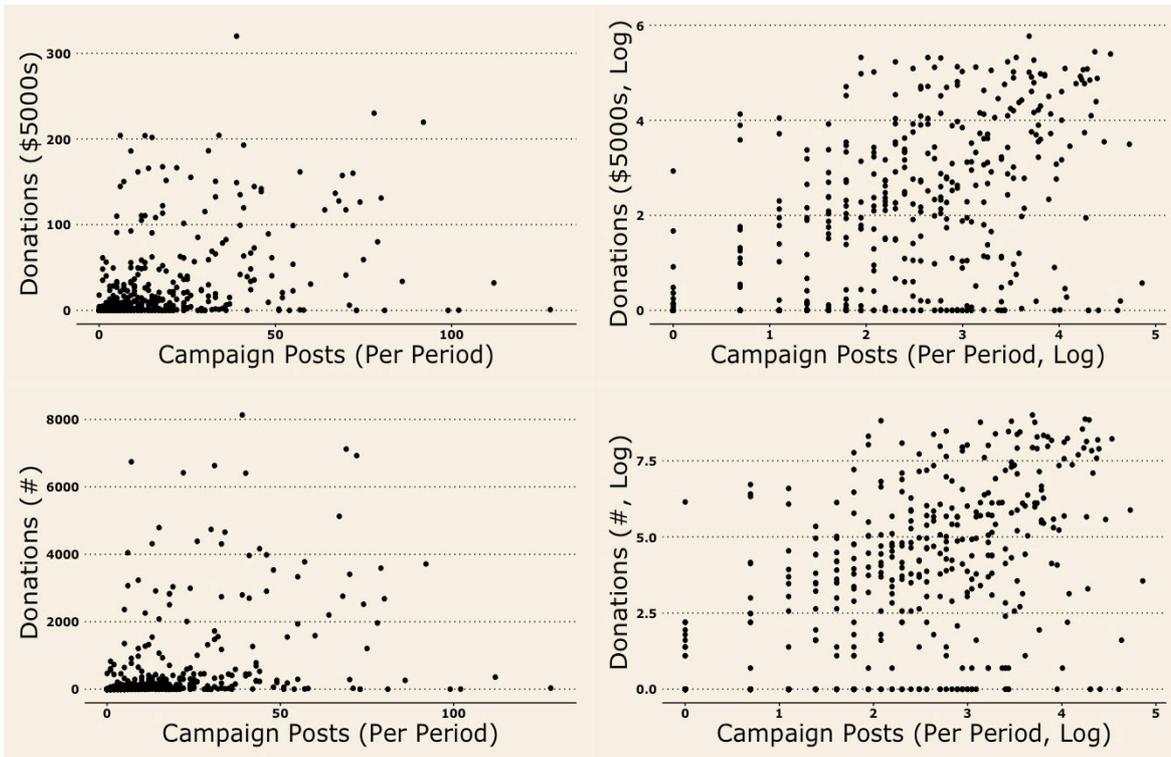

**Figure SI.22. Scatterplot of Campaign-Related Posts versus Donations, Linear (Left) and Log-Transformed (Right)**. *The relationship is not linear when in lin-lin form, suggesting the need for log-transformation.*

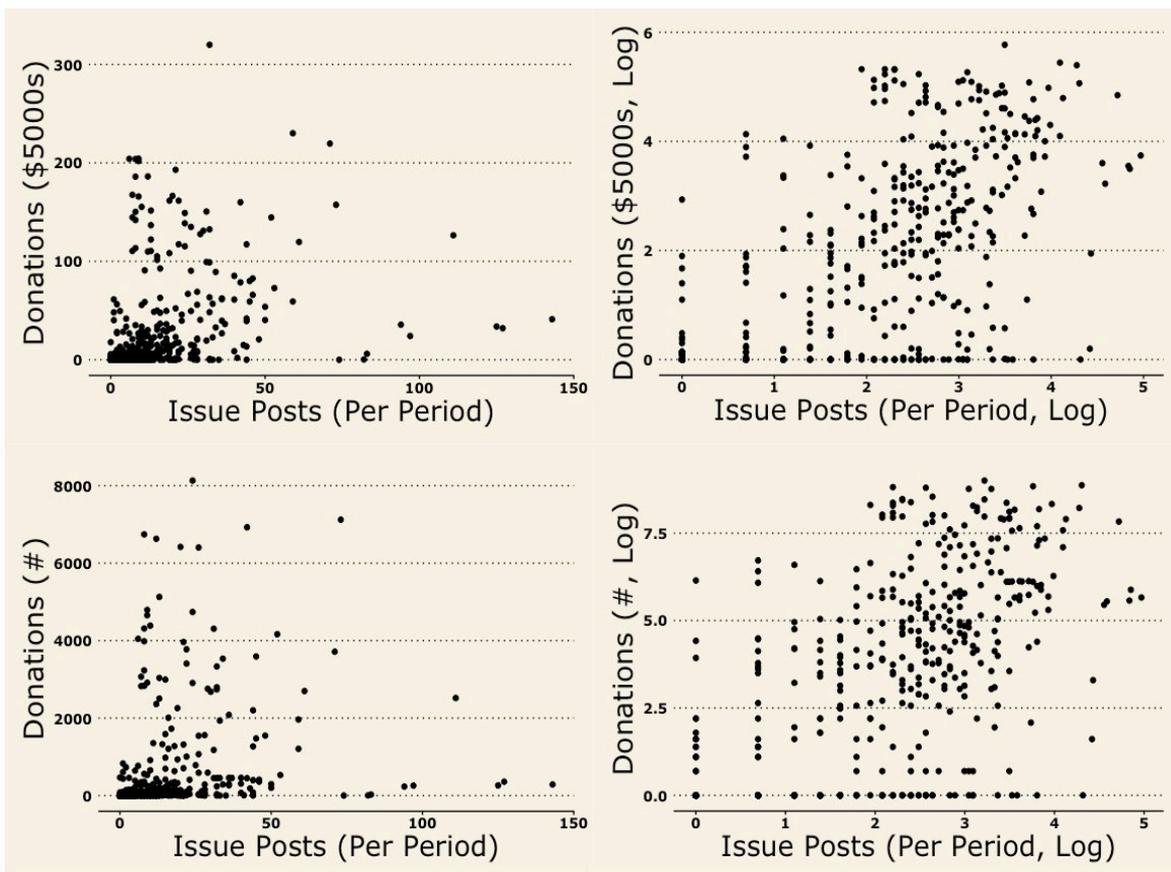

**Figure SI.23. Scatterplot of Issue-Related Posts versus Donations, Linear (Left) and Log-Transformed (Right)**. *The relationship is not linear when in lin-lin form, suggesting the need for log-transformation.*

# Regression Model Diagnostics

## Residual versus Fitted Plots

Residual versus Fitted plots reveal if residuals have non-linear patterns. Residuals for models based on log-transformed data appear to be relatively evenly spread with no clear patterns, which indicates a linear relationship.

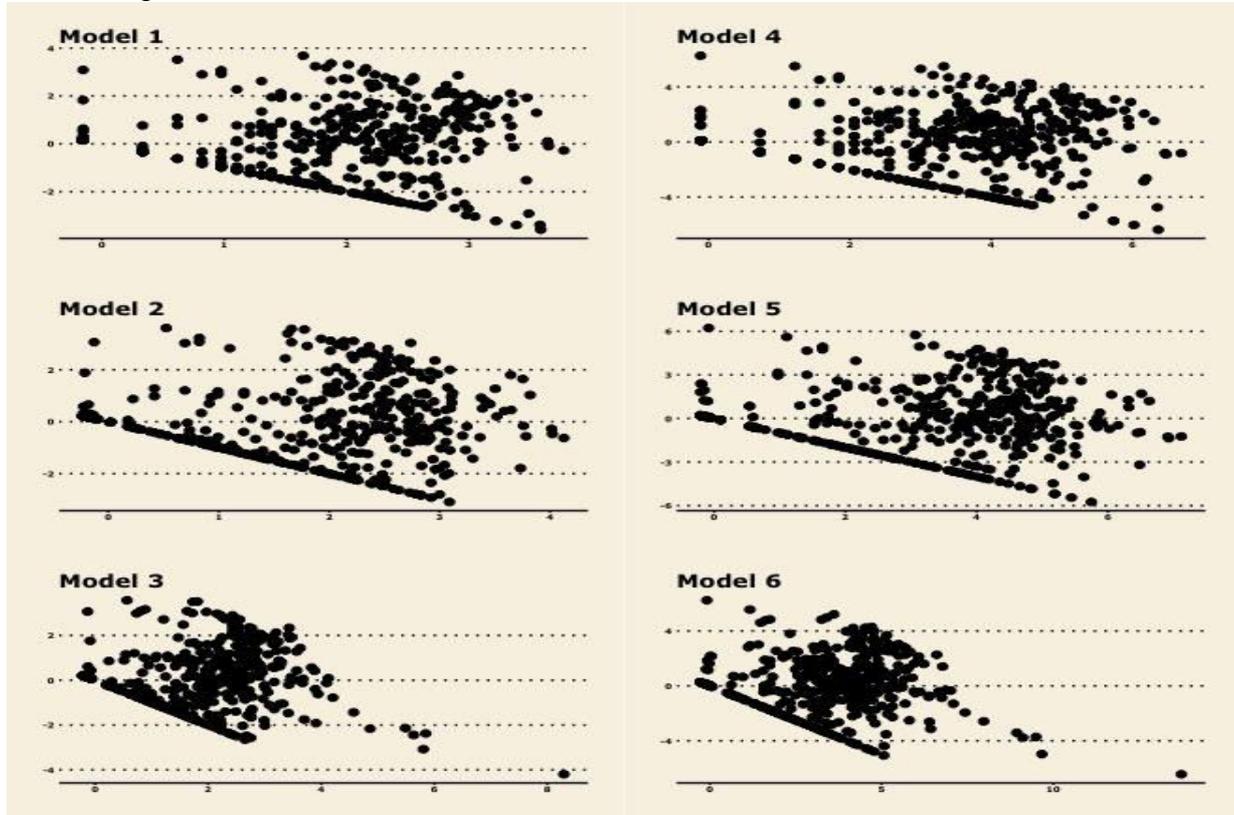

**Figure SI.24. Residuals versus Fitted, with IV=Post Volume.**

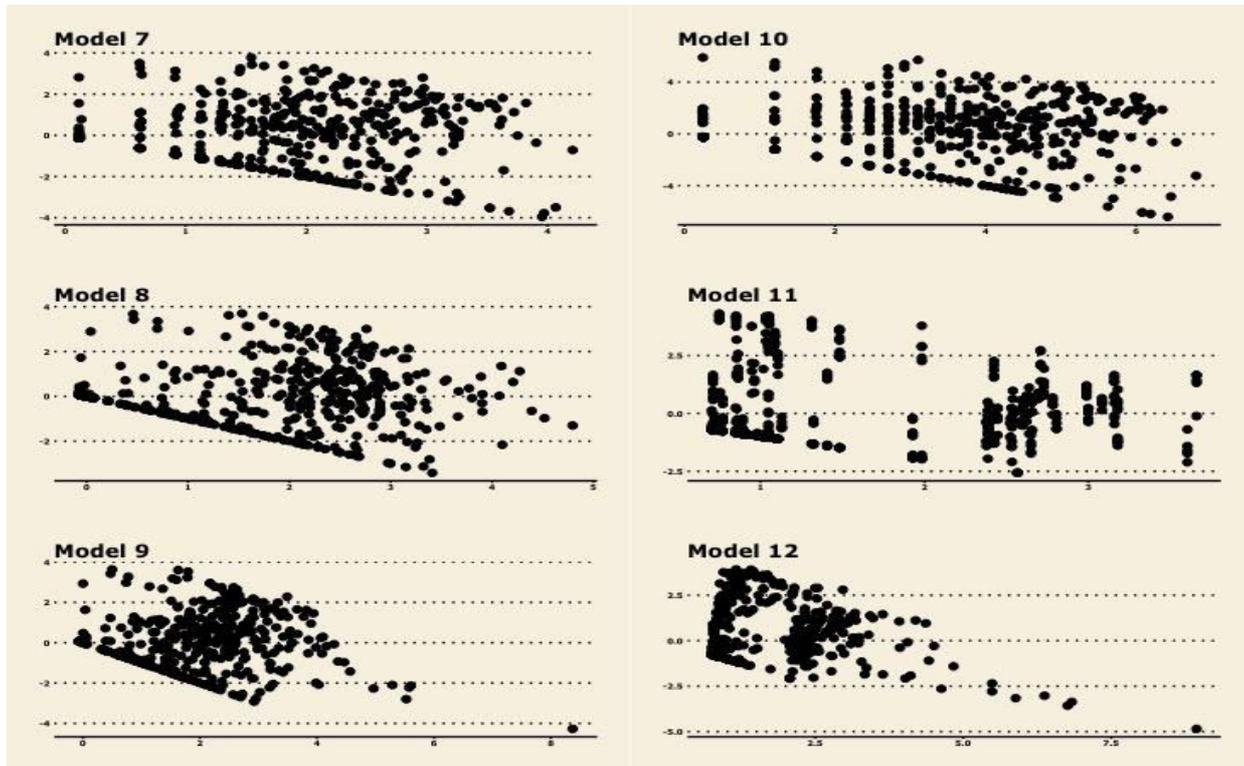

**Figure SI.25. Residuals versus Fitted, with IV=Campaign-Related Posts.**

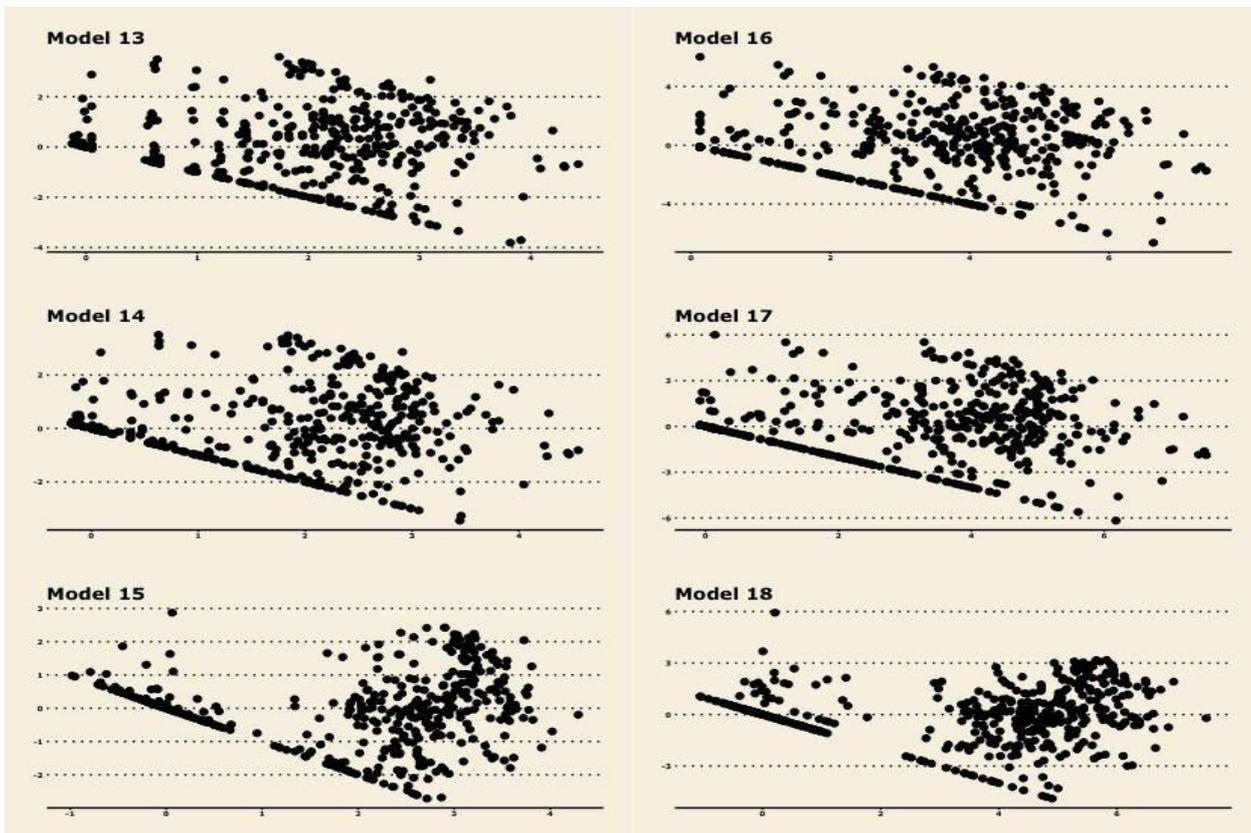

**Figure SI.26. Residuals versus Fitted, with IV=Issue-Related Posts.**

# Normal Q-Q Plots

Normal Q-Q plots based on log-transformed data show the residuals are normally-distributed, as the points of the plot lie roughly on the straight diagonal lines.

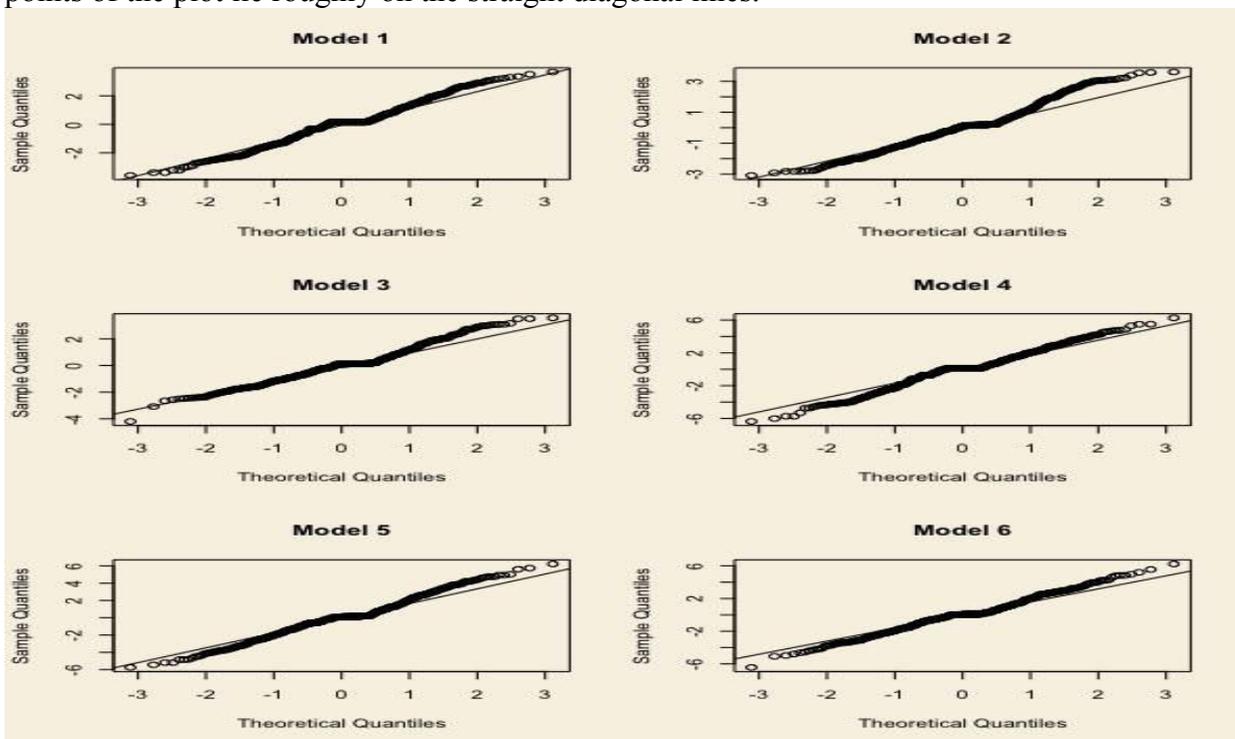

**Figure SI.27. QQ-Normal Plots, with IV=Post Volume.**

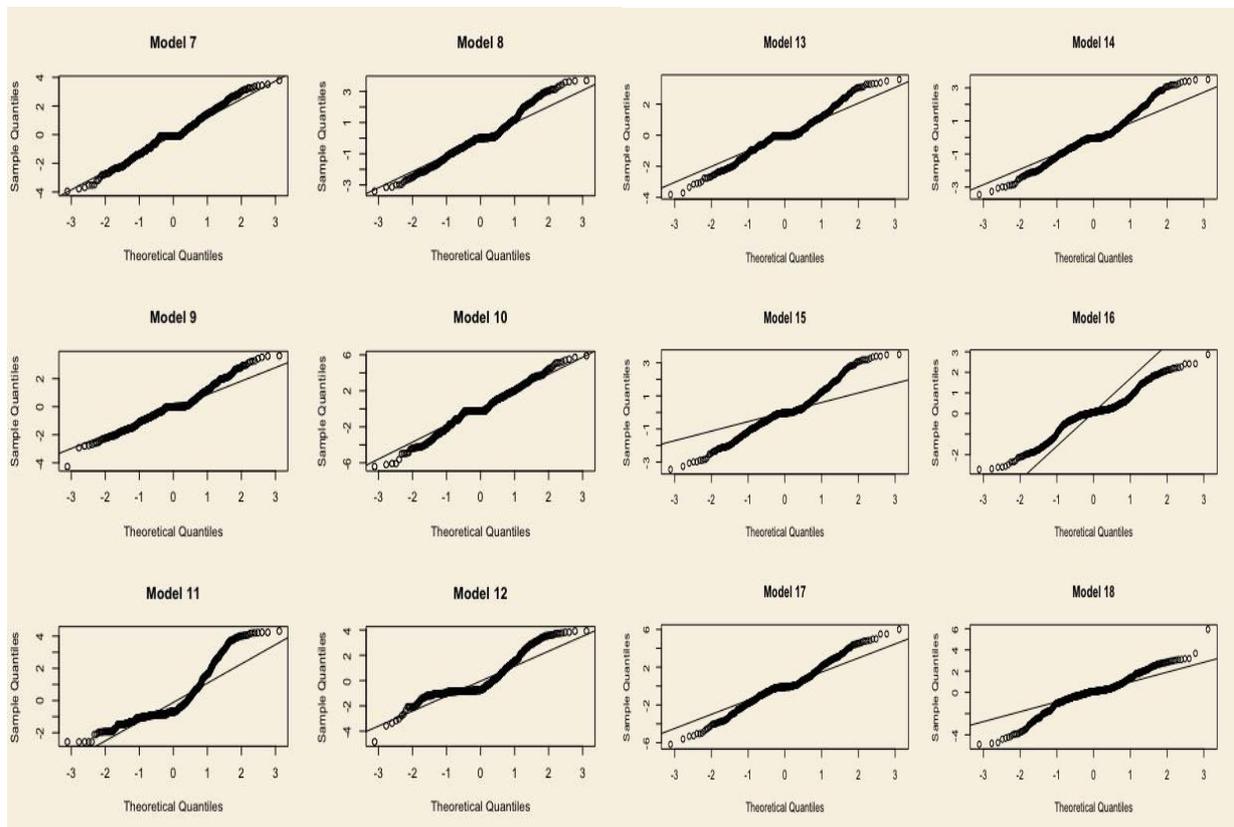

**Figure SI.28.** QQ-Normal Plots, with IV=Campaign-Related Posts (Models 7-12) and IV=Issue-Related Posts.

## *Correlation between Itemized and Unitemized Sums*

To determine whether the unitemized sums can be taken as a proxy for the full value received by candidates, we examine the relationship between unitemized and itemized contributions over the 2016 cycle for each candidate with a campaign account. This involves collecting data on the itemized and unitemized receipts manually through the FEC website: https://www.fec.gov/data/. There are a total of 78 candidates with both sums listed for the 2016 cycle. The two variables are highly positively correlated, with a Pearson's correlation coefficient of 0.71.

# CODE

In the interest of replicability, code used to access and analyze all data involved in this research may be found here: https://figshare.com/s/ff85b0a22fe470de5e80.